\documentclass[aps,prx,twocolumn]{revtex4-2}

\usepackage{amsmath,amssymb,graphicx,color,bm,epstopdf}
\usepackage{bbold}
\usepackage{braket}
\usepackage{enumitem}
\usepackage{orcidlink}
\usepackage{tikz}
\usetikzlibrary{shapes,arrows,decorations.pathmorphing}
\usepackage{dsfont}
\usepackage{stmaryrd}
\usepackage{mathtools}

\newcommand\norm[1]{\lVert#1\rVert}
\newlist{UR}{enumerate}{1}
\setlist[UR]{label=[M\arabic*]}
\usepackage[resetlabels,labeled]{multibib}
\newcites{S}{Supplementary References}

\begin{document}

\title{Quantum Control of Radical Pair Dynamics beyond Time-Local Optimization}
\author{Farhan T.\ Chowdhury\,\orcidlink{0000-0001-8229-2374}}
\author{Matt C.\ J.\ Denton\,\orcidlink{0009-0009-0246-3387}}
\author{Daniel C.\ Bonser\,\orcidlink{0009-0002-0749-9273}}
\author{Daniel R.\ Kattnig\,\orcidlink{0000-0003-4236-2627}}
\affiliation{Department of Physics and Astronomy, University of Exeter, Stocker Road, Exeter EX4 4QL, UK}

\begin{abstract}

We realize arbitrary waveform-based control of spin-selective recombination reactions of radical pairs in the low magnetic field regime. To this end, we extend the Gradient Ascent Pulse Engineering (GRAPE) paradigm to allow for optimizing reaction yields. This overcomes drawbacks of previously suggested time-local optimization approaches for the reaction control of radical pairs, which were limited to high biasing fields. We demonstrate how efficient time-global optimization of the recombination yields can be realized by gradient based methods augmented by time-blocking, sparse sampling of the yield, and evaluation of the central single time-step propagators and their Fr\'echet derivatives using iterated Trotter-Suzuki splittings. Results are shown for both a toy model, previously used to demonstrate coherent control of radical pair reactions in the simpler high-field scenario, and furthermore for a realistic exciplex-forming donor-acceptor system comprising 16 nuclear spins. This raises prospects for the spin-control of actual radical pair systems in ambient magnetic fields, by suppressing or boosting radical reaction yields using purpose-specific radio-frequency waveforms, paving the way for reaction-yield-dependent quantum magnetometry and potentially applications of quantum control to biochemical radical pair reactions. We demonstrate the latter aspect for two radical pairs implicated in quantum biology.
\end{abstract}

\maketitle

\section{Introduction} 

The rapidly evolving field of optimal quantum control (OQC) seeks to manipulate phenomena at the quantum scale by devising and implementing perturbations, typically in the form of electromagnetic pulses, to steer a given quantum system to a desired target. Fuelled by demonstrable successes in nuclear magnetic resonance (NMR) pulse engineering \cite{cont22, kosloff86, chemOC} and the control of ultra-fast excited-state reactions by laser fields \cite{fmQC}, pioneering developments in ``coherent control" were first conceived in chemistry. Today OQC has developed into a mature discipline that is central to modern quantum technologies associated with the second quantum revolution \cite{jonSQC}, with numerous applications across quantum information processing and quantum metrology \cite{cont22}. Following on from an earlier proposal for controlling the dynamics of radical pair reactions through modulating the exchange interaction via an optically switchable bridging group \cite{guerreschi2013,vitalis2017}, a recent suggestion extends coherent control to radical pair reactions mediated through radio-frequency magnetic fields \cite{mae}, thereby putting a renewed spotlight on the quantum control of spin-chemical reactions. In the present work we go on to demonstrate the implementation of a Gradient Ascent Pulse Engineering (GRAPE)-inspired approach, made computationally feasible via novel algorithmic modifications, towards realizing open-loop control of singlet recombination yields for radical pair dynamics in weak magnetic fields.

Reactions involving the recombination of radical pair intermediates are well-known to depend on spin degrees of freedom and their intrinsic quantum dynamics. Specifically, the electronic singlet and triplet states associated with the unpaired electrons for each radical can undergo coherent inter-conversion as a consequence of symmetry-breaking interactions—in particular, the hyperfine interactions with surrounding magnetic nuclei. Quite recently, quantum beats reflecting the coherent singlet-triplet interconversion in radical pairs have been directly revealed through pump-push spectroscopy \cite{Mims2021}. In the case of singlet and triplet states exhibiting differential chemical reactivity, which is usually the case since radical pair recombination preserves spin-multiplicity in the absence of strong spin-orbit coupling, the spin dynamics is reflected in the reaction yields realized via the singlet and triplet channels. By coupling to magnetic fields via the Zeeman interaction, radical pairs furthermore acquire sensitivity to static and oscillatory applied magnetic fields. The former is central to various magnetic field effects (MFEs) associated with radical pair reactions, while the latter are pertinent to the prospect of radical pair reaction control through radio-frequency magnetic fields. Studies of effects of oscillatory magnetic fields on radical pair reactions have been realized for chemical model systems and play a discriminatory role in identifying magnetosensitive radical pair reactions in spin-biological systems \cite{pophof23, pophof23b}. 

While a majority of model studies on oscillatory magnetic field effects have applied monochromatic radio-frequency magnetic fields in the presence of a strong biasing field \cite{wasi83}, often referred to as reaction yield detected magnetic resonance (RYDMR), some studies have been realized in a weak static field, in particular for exciplex-forming systems \cite{rydmrA, rydmrB, rydmrC, exci, mani22}. In such systems, consisting of an electronically excited donor-acceptor complex characterized by partial charge transfer and photo-emissivity \cite{Rugg2019}, the radical pair is in equilibrium with or can populate an exciplex state. In this way, radical pair dynamics become accessible/measurable via the exciplex emission, i.e.\ the recombination yield is proportional to the fluorescence emission under steady-state conditions \cite{Wedge2013}. Radical pair mediated mechanisms, or variations thereof \cite{kt21}, are also hypothesized to underpin a compass sense in various animals, including migratory songbirds \cite{hore16}. In this case, the magnetosensitive radical pair is thought to originate from a photo-induced electron transfer reaction in the flavo-protein cryptochrome. Even though the exact nature of the underlying radical pair has remained unclear, the fact that this compass sense is interruptible by weak monochromatic \cite{ritz09} or broad-band radio-frequency \cite{engels14} magnetic fields strongly supports an underlying spin mediated mechanism. These observations support the prospect of controlling chemical reactions involving radical pairs, possibly in the near future, in the spin-biological context, and the challenging low-field regime, provided that an efficient approach to pulse engineering can be developed, as we set out to do here.

Recent work \cite{mae, mae2} suggested the control of radical pair reactions based on model calculations for a prototypical spin system. They focus on the traditional RYDMR scenario, for which the applied electromagnetic field is in resonance with the electronic Zeeman splitting produced by a strong static field. This perturbs the singlet-triplet interconversion of the radical pair and thus affects the yield of geminate recombination. Their control approach, building upon the theoretical approach developed by Sugawara \cite{sugi03}, propagates the spin density operator whereby at every moment the control amplitude is chosen such that an optimization criterion is bound to not decrease. Since the algorithm optimizes the controls in a time-local fashion, it is dubbed as \emph{local optimization}. By construction, the employed algorithm allows optimizing the trajectory of an observable commuting with the time-independent dynamics generator, i.e.\ the drift Hamiltonian, or alternatively, an arbitrary observable at \emph{one} chosen moment in time (via back-propagation). While the combined population of the singlet ($S$) and one of the triplet states ($T_0$) is hence controllable over time in the high-field limit, the more relevant singlet population falls in the second category. Through our present work, we overcome ostensible drawbacks of the time-local optimization approach by permitting time-global optimization that simultaneously optimizes the controls for all times and by focusing on the singlet recombination yield, i.e.\ the actual experimental observable (rather than $S$ and $T_0$-populations over time or the singlet probability at a moment of time). We achieve this through using piece-wise constant control amplitudes and gradient information, by building upon Gradient Ascent Pulse Engineering (GRAPE) \cite{oqc05}. To realize quantum control of radical pair dynamics in a weak static magnetic field, these developments are critical.

Our approach, even in the framework of controls derived for prototypical models not explicitly considering weakly coupled nuclear spins and spin relaxation processes, promises robustness in applications to more realistic system descriptions. This opens avenues for an OQC based model-tomography approach to distinguish between candidate radical pairs in systems where magnetosensitivity is phenomenologically observable, but the underpinning mechanisms remain opaque. Enabling the discriminatory shaping of magnetosensitive responses could prove to be particularly fruitful for pinning down key facilitating mechanisms, with applications ranging from designing organic spintronic materials \cite{spintron}, to enhancing the imaging of magnetic nanostructures \cite{lee2011mapping}. Further potential avenues for exploration are magnetic field measurements, e.g.\ via delayed fluorescence-based organic light emitting diodes with large magneto-electroluminescence \cite{acsMFS}, or for medical and biological applications \cite{simon2022}, where selected pathways could potentially be controlled, possibly to endpoints not reachable via static magnetic field exposure alone.

We deviate from traditional GRAPE by optimizing reaction yields (rather than a fidelity measure defined at a given moment in time), accounting for asymmetric radical pair recombination, which gives rise to non-unitary propagators, using exact gradients rather than approximate ones, and, following de Fouquieres et al.\ \cite{qLbfgs}, by using curvature information of the loss function (rather than steepest decent/ascent). To overcome a major objection to GRAPE-based reaction control of radical pairs, namely high computational demands, as also voiced by Masuzawa et al.\ \cite{mae}, we suggest a number of practical optimizations such as a block-optimization scheme and sparse sampling of the reaction yield to improve the efficiency of the approach while retaining adequate control fidelity. With these improvements, realistically large radical pair systems can be yield-controlled in weak magnetic fields. We demonstrate the success of the approach for a large spin system in an exciplex-forming complex and further provide examples motivated by quantum biology.

\section{Theory} 

\subsection{Radical Pair Spin Dynamics}
\begin{figure} [h]
\includegraphics[scale=0.84]{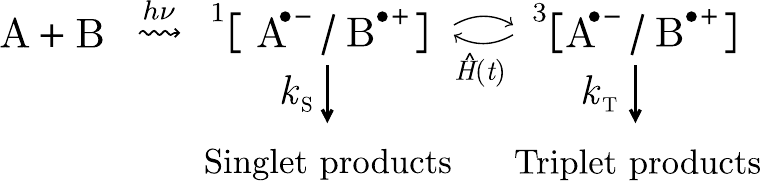}
\caption{Reaction Scheme.}
\label{fig:reaction}
\vspace{-1em}
\end{figure}

We consider a system comprising radicals A$^{\bullet-}$ and B$^{\bullet+}$ subject to singlet-triplet interconversion, undergoing spin-selective recombination reactions as per Fig.~\ref{fig:reaction}.

\noindent The corresponding spin dynamics is described in terms of the time-dependent spin density matrix $\hat{\rho}(t)$ by
\begin{align}
\frac{d\hat{\rho}(t)}{dt} =-i[\hat{H}(t),{\hat{\rho}(t)}] - \{\hat{K}, \hat{\rho}(t)\}, \label{eq:lindblad}
\end{align}
with $\hat{H}(t)$ denoting the spin Hamiltonian and 
\begin{align}
\hat{K} = \frac{k_{\rm S}}{2}\hat{P}_{\rm S} + \frac{k_{\rm T}}{2}\hat{P}_{\rm T}
\end{align}
accounting for radical pair recombination. Here, $k_{\rm S}$ and $k_{\rm T}$ denote the reaction rate constants in the singlet and triplet state, respectively, and $\hat{P}_{\rm S}, \hat{P}_{\rm T}$ are the projection operators onto the singlet and triplet states. We introduce an effective Hamiltonian given by
\begin{align}
\hat{A}(t) = \hat{H}(t) - i\hat{K} ,
\end{align} 
allowing us to reformulate eq.~(\ref{eq:lindblad}) as
\begin{align}
\frac{\partial \hat{\rho}(t)}{\partial t} = -i\llbracket \hat{A}(t),\hat{\rho}(t) \rrbracket \nonumber \\
= -i(\hat{A}(t)\hat{\rho}(t) -\hat{\rho}(t) \hat{A}(T)^{\dagger}), \label{eq:truncated}
\end{align}
where $\llbracket A,B \rrbracket= AB -(AB)^{\dagger} = AB - BA^{\dagger}$ $\forall$ $B=B^{\dagger}$. The formal solution to eq.~(\ref{eq:truncated}) is given by
\begin{align}
\hat{\rho}(t_n) = \hat{U}(t_n,0)\hat{\rho}(0)\hat{U}^{\dagger}(t_n,0)
\end{align}
where the time evolution operator in terms of $\hat{A}(t)$ is  
\begin{align}
    \hat{U}(t,0) = \mathcal{T} \mathrm{exp}\left[-i\int_{0}^{t} \hat{A}(\tau)\mathrm{d}\tau\right]. \label{eq:unitary}
\end{align}
Here, $\mathcal{T}$ denotes the time ordering operator, that reorders products of time-dependent operators such that their time arguments decrease going from left to right. 

The singlet channel recombination yield is given by
\begin{align}
Y_{\rm S}=k_{\rm S}\int_{0}^{\infty} p_{\rm S}(t) dt
\label{eq:yield}
\end{align}
where $p_{\rm S}(t)=Tr[\hat{P}_{\rm S}\hat{\rho}(t)]$ is the survived singlet probability of the radical pair. Note that eq.~(\ref{eq:truncated}) is of truncated Lindblad form (by introducing shelving states, an equivalent formulation in traditional Lindblad form is also possible \cite{gauger11}). Thus, the essentially open quantum system description of radical pair spin dynamics in the presence of spin-selective recombination eq.~(\ref{eq:lindblad}) can be treated in terms of the dynamics of a closed system, albeit with a non-Hermitian dynamics generator, non-unitary evolution operators and non-conserved trace.  

Here, we consider radical pairs for which the static part of the Hamiltonian, $H_0$, is of the form
\begin{align}
    \hat{H}_0 = \hat{H}_A + \hat{H}_B + \hat{H}_{AB}
\end{align}
where $\hat{H}_i, i \in \{A,B\}$ is local to radical $i$ and comprises the Zeeman interaction with the static magnetic field and isotropic hyperfine couplings with surrounding nuclear spins. $\hat{H}_{AB}$ accounts for inter-radical couplings in the form of the exchange interaction. Specifically, each radical is described in terms of corresponding nuclear and electron spin angular momentum operators $\hat{\bf I}_{ik}$ and $\hat{\bf S}_i$ by
\begin{align}
    \hat{H}_i = {\omega}_i\hat{S}_{iz}+\sum_{k=1}^{n_i} a_{ik}\hat{\bf I}_{ik}\cdot\hat{\bf S}_i. 
\end{align}
and
\begin{align}
    \hat{H}_{AB} = -j_{ex}\left(2\hat{\bf S}_A\cdot\hat{\bf S}_B + \frac{1}{2}\right)
\end{align}
where ${\omega}_i=-\gamma_iB$, with $\gamma_i$ denoting the gyromagnetic ratio of the electron in radical $i$ and $B$ the applied magnetic field, $a_{ik}$ is the isotropic hyperfine coupling constant between electron spin $i$ and the $k$-th nuclear spin (out of total $n_i$), and the exchange coupling constant is denoted by $j_{ex}$. The initial density operator is assumed as
\begin{align}
  \hat{\rho}(0) = \frac{1}{Z_1Z_2}\hat{P}_{\rm S},  
\end{align}
where $Z_i=\prod_{k=1}^{n_i} (2I_{ik}+1)$ is the total number of nuclear spin states in radical $i$ and
\begin{align}
\hat{P}_{\rm S}= \frac{1}{4}-\hat{\bf S}_1\cdot\hat{\bf S}_2, 
\end{align}
is the singlet projection operator. In the context of spin-biological systems, we also consider directional magnetic field effects of systems with anisotropic hyperfine interactions. In this case, $\hat{H}_i$ is of the form
\begin{align}
    \hat{H}_i = {\omega}_i {\bf n}(\vartheta,\varphi)\cdot\hat{\bf S}_i+\sum_{k=1}^{n_i} \hat{\bf I}_{ik}\cdot {\bf A}_{ik}\cdot\hat{\bf S}_i. 
\end{align}
where ${\bf n}(\vartheta,\varphi)$ is an unit vector with polar angle $\vartheta$ and azimuth $\varphi$ defining the magnetic field direction and the ${\bf A}_{ik}$s are the hyperfine tensors.

\subsection{Gradient-based coherent control}

We here outline pertinent features and the required extensions for our GRAPE-inspired approach. Borrowing its basic set up from GRAPE, our approach entails discretizing controls into time slices that are improved in a concurrent-update scheme, i.e.\ involving controls for all time points (thus ``time-global"), with each update determined only by information from the previous iteration, so they can be evaluated at each point independently. Specifically in GRAPE, the gradient of the fidelity (usually defined at a predefined final time) with respect to all control variables is calculated (or rather approximated) in each iteration and a step is made in the variable space in the direction of steepest ascent (or descent). For quantum systems controllable via convex optimization, first-order gradient information can sufficiently guarantee that the global maximum at the top of the control landscape can be located from any given direction, with the extremes of a given control landscape corresponding to either perfect control or no-control \cite{Rabitz2004}. 

Without loss of generality, we can describe a given quantum system undergoing spin dynamics under coherent control with
\begin{align}
\hat{H}(t) = \hat{H}_{0} + \sum_{l=1}^{M} {u}_{l}(t) \hat{V}_{l}, \label{eq:control}
\end{align} 
where $\hat{H}_0$ is the drift Hamiltonian describing the time-independent part of the system, i.e.\ the intrinsic radical pair dynamics in the static bias field, as introduced above, ${u}_{l}(t)$ are the control amplitudes and $\hat{V}_l$ is the set of control Hamiltonians that couple the controls to the system, for example via the Zeeman interaction. We control parameters of only the coherent evolution of the open system dynamics of the radical pair undergoing spin selective recombination reactions. In principle, the recombination could also be directly controlled. However, direct control of recombination is less amenable to experimental manipulation and therefore not pursued here.

The goal of OQC is to optimize some control objective functional $G[\{{u}_{l}(t)\}]$, with a control target subject to physical constraints and other criteria. For our purposes, we focus on minimizing or maximizing the singlet recombination yield of the radical pair, i.e.\ $G=Y_{\rm S}$. However, the approach is easily generalized to focus on other reaction outcomes, e.g.\ accumulated nuclear polarization. Significant prior work has been concentrated on optimizing $G$ subject to the assumption that the values taken by the controls over time may be parameterized by piece-wise constant control amplitudes in the time domain \cite{machnes11}. This paradigm, which is central to GRAPE as the first widely applied quantum control algorithm \cite{oqc05}, is well aligned with the practical aspect of controlling radical pairs, as standard arbitrary waveform generator (AWG) based control schemes represent pulse inputs as piece-wise constant functions. We thus discretize the time axis, $t_0 = 0 < t_1 < t_2 < \cdots < t_N = t_{max}$ and we set 
\begin{align}
    \hat{H}(t) \approx \hat{H}_0 + \sum_{l=1}^{M}\sum_{n=1}^{N} u_{l,n}\chi_n(t) \hat{V_l}
\end{align}
with constants ${u}_{l,n}$ and the indicator function $\chi_n(t)$ defined such that  $\chi_n(t)=\begin{cases}
        1 & \text{for $t_{n-1}\leq t<t_n$}\\
        0 & \text{otherwise}.
        \end{cases}.$

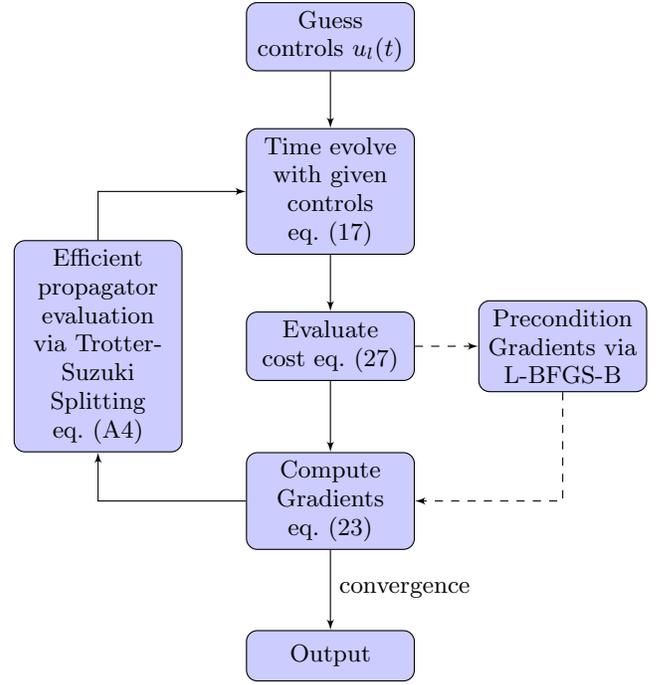
\begin{figure}
    \centering
        \resizebox{\linewidth}{!}{
\begin{tikzpicture}[node distance = 2cm, auto]
\tikzstyle{block} = [rectangle, draw, fill=blue!20, 
    text width=6em, text centered, rounded corners, node distance=2cm, minimum height=2em]
\tikzstyle{line} = [draw, -latex']
\tikzstyle{block2} = [rectangle, draw, fill=blue!20, 
    text width=6em, text centered, rounded corners, node distance=3cm, minimum height=2em]
\tikzstyle{line} = [draw, -latex']
    \node [block] (init) {Guess controls ${u}_{l}(t)$};
    \node [block, below of=init] (identify) {Time evolve with given controls eq.~(\ref{eq:evolveT})};
    \node [block, below of=identify] (evaluate) {Evaluate cost eq.~(\ref{eq:approxG})};
    \node [block2, right of=evaluate] (system) {Precondition Gradients via L-BFGS-B};
    \node [block, left of=evaluate, node distance=3cm] (update) {Efficient propagator evaluation via Trotter-Suzuki Splitting eq.~(\ref{eq:expTrot})};
    \node [block, below of=evaluate] (decide) {Compute Gradients eq.~(\ref{eq:gradients})};
    \node [block, below of=decide] (stop) {Output};
    \path [line] (init) -- (identify);
    \path [line] (identify) -- (evaluate);
    \path [line] (evaluate) -- (decide);
    \path [line] (decide) -| node [near start] {} (update);x
    \path [line] (update) |- (identify);
    \path [line] (decide) -- node {convergence}(stop);
    \path [line,dashed] (evaluate) -- (system);
    \path [line,dashed] (system) |- (decide);
\end{tikzpicture}
	}
    \caption {A simplified flowchart for optimizing radical pair reaction yields. Initially, the control variables are guessed, and the system is time evolved. The fidelity is then computed in terms of an objective function, and we get an output where convergence is achieved. Otherwise, the gradient with respect to control variables is computed and utilized, together with iteratively accumulated curvature information, to update the controls, using which the system is again time evolved and the process repeats.}
\label{fig:flow}
\vspace{-1em}
\end{figure}            

The corresponding propagators $\hat{U}(t_n,0)$ are discretized on the temporal grid such that
\begin{align}
\hat{U}(t_n,0) = \mathcal{T} \prod_{i=1}^{n} \hat{U}_i = \hat{U}_n \cdots \hat{U}_1
\end{align}
where
\begin{align}
\hat{U}_i = e^{-i \hat{A}_i \Delta t_i}  \label{eq:evolveT}
\end{align}
with $\Delta t_i=t_i - t_{i-1}$ and $\hat{A}_i=\hat{A}(t_{i-1})$. The gradients of $G$ depend on the propagator derivatives, evaluated as
\begin{align}
\frac{\partial \hat{U}_{i}}{\partial {u}_{l,j}} = \delta_{i,j} L_{i,l} = \delta_{i,j} L(i \hat{A}_{i} \Delta t_i , i\hat{V_l}\Delta t_{i})
\end{align}
where $L(X,E)$ is the Fr\'echet derivative of the matrix exponent defined, for every $E$, by 
\begin{align}
e^{(X+E)} - e^X = L(X,E) + o(\norm{E}).
\label{eq:frechet}
\end{align}
More explicitly, we have \cite{wilcox67}
\begin{align}
L(X,E)= \int_0^1{e^{X(1-s)}} E e^{Xs} ds\nonumber \\
= \sum_{n=1}^{\infty} \frac{1}{n!} \sum_{j=1}^{n} X^{j-1} EX^{n-j},
\end{align}
which also allows us to demonstrate that
\begin{align}
\frac{\partial \hat{U}_{i}^{\dagger}}{\partial {u}_{l,j}} = \delta_{i,j} L(+i \hat{A}_{i}^{\dagger}\Delta t_i ; +i\hat{V_l}\Delta t_{i}) \nonumber \\
= \delta_{i,j} L^{\dagger}(-i \hat{A}_{i}\Delta t_i ; -i\hat{V_l}\Delta t_{i})\nonumber \\
= \delta_{i,j} L_{i,l}^{\dagger}
\end{align}

Combining the above equations, the gradient of the survived singlet probability, 
\begin{align}
p_{\rm S}(t_n) = Tr[\hat{P}_{\rm S} \hat{U}(t_n,0)\hat{\rho}(0)\hat{U}^{\dagger}(t_n,0)],
\label{eq:sp} 
\end{align}
for $k \leq n$, is thus
\begin{align}
\frac{\partial p_{\rm S}}{\partial {u}_{l,k}} = -i\textrm{Tr}(\hat{P}_{\rm S}\hat{U}_{+} \llbracket  \hat{W}_{k,l}, \hat{U}_{-}\hat{\rho}_{0}\hat{U}_{-}^{\dagger} \rrbracket \hat{U}_{+}^{\dagger}) \nonumber \\
= -i\textrm{Tr}(\hat{U}_{+}^{\dagger}\hat{P}_{\rm S} \hat{U}_+ \llbracket \hat{W}_{k,l}, \hat{U}_{-}\hat{\rho}_{0}\hat{U}^{\dagger}_{-} \rrbracket)\nonumber \\
= 2\Im[\textrm{Tr}(\hat{U}_{+}^{\dagger}\hat{P}_{\rm S} \hat{U}_+ \hat{W}_{k,l} \hat{U}_{-}\hat{\rho}_{0}\hat{U}^{\dagger}_{-})]\nonumber \\
= 2\Im[\textrm{Tr}(\hat{P}_{\rm S} \hat{U}_+ \hat{W}_{k,l} \hat{U}_{-}\hat{\rho}_{0}\hat{U}^{\dagger})]   \label{eq:gradients}
\end{align}
where $\hat{W}_{i,l} = i\hat{U}_{i}^{-1} L_{i,l}$ and
\begin{align}
\hat{U}_{+} = \mathcal{T} \prod_{m=k}^{n} \hat{U}_m\\
\hat{U}_{-} = \mathcal{T} \prod_{m=1}^{k-1} \hat{U}_m.
\end{align}
For $k > n$, the gradient obviously vanishes. In the original formulation of GRAPE, a first-order approximation of $L_{k,l}$ was used \cite{oqc05}, which is obtained with $\hat{W}_{i,l} = \hat{V_l}$. For our approach, we use the full Fr\'echet derivative. Finally, this allows us to approximate the singlet yield by 
\begin{align}
Y_{\rm S} = k_{\rm S} \int_{0}^{\infty} p_{\rm S}(t) dt\nonumber \\ 
\approx k_{\rm S} \sum_{n=0}^{N} {w}_n p_{\rm S} (t_n) 
\label{eq:approx} 
\end{align}
with ${w}_n$ denoting suitably chose quadrature weights. The gradient of the singlet yield is then obtained as
\begin{align}
\frac{\partial Y_{\rm S}}{\partial {u}_{l,k}} \approx k_{\rm S}\sum_{n=k}^{N} {w}_n\frac{\partial p_{\rm S}}{\partial {u}_{l,k}}.
\label{eq:approxG} 
\end{align}

\section{Results}

\begin{figure}
    \centering
    \includegraphics[width=0.8\columnwidth]{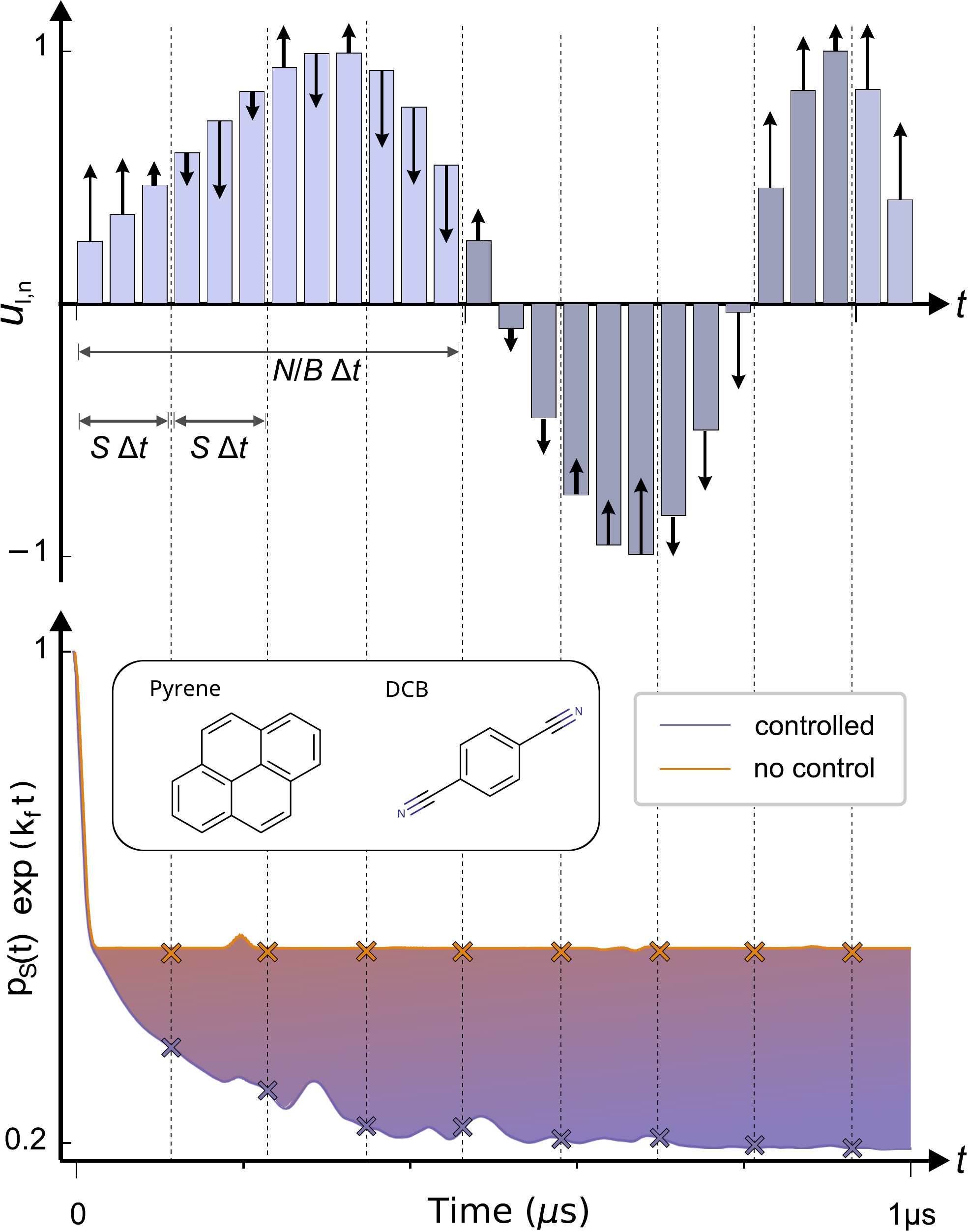}
    \caption{Schematic illustration of the time-blocking and sparse-sampling approach used to reduce the computational expense of gradient-based optimization of the radical pair recombination yield. The control amplitudes ${u}_{\text{l,n}}$ are simultaneously updated, whereby $B$ disjoint blocks of $\frac{N}{B}$ time-steps each are considered in succession, with the initial state taken to be that of the system subject to the preceding optimized controls. A sparsely sampled approximation of the reaction yields based on $\frac{N}{BS}$ samples per block suffices to effectively optimize the reaction yield, in particular for larger spin systems for which the singlet probability is barely oscillatory. The bottom panel illustrates the unperturbed (orange solid line) and minimized (blue) singlet probability for Py/DCB, an 18 spin exciplex system in a static biasing field of $1\mskip3mu$mT. The RF control field of $0.1\mskip3mu$mT amplitude comprised $1024$ control pulses of $\mskip3mu1$ns width, the amplitude of which was optimized such that the singlet yield approximately calculated from $256$ samples of the singlet probability was minimized assuming a recombination rate of $k=1\mskip3mu\mu$s$^-1$. The  Py/DCB system is discussed in more detail in the Results section.}
\label{fig:sketch}
\end{figure}

The control of radical pair dynamics using time-global optimization methods is expected to be computationally costly, which is why the possibility of adapting GRAPE was dismissed in favour of time-local optimization in earlier attempts \cite{mae, mae2} at realizing radical pair-reaction control. The computational demand is particularly large if, as we do here (and unlike \cite{mae, mae2}), the goal is to optimize yields, which are time-integral quantities (cf.\ eq.~(\ref{eq:yield})), unlike the typical fidelity measures in quantum control applications pertaining to the implementation of a state transfer or a unitary operation for a fixed time $t_{\rm max}$. Nevertheless, by leveraging various algorithmic modifications, we succeed in demonstrating the feasibility of a time-global approach for controlling the dynamics of radical pair spin systems of realistic complexity.

\subsection{Algorithmic efficiency improvements} 

We require computing $N$ single-time-step propagators alongside their Fr\'echet derivatives with respect to $L$ control Hamiltonians, and scores of repeated matrix multiplications to evaluate the time-dependent singlet-probabilities and associated gradients, which sum to the recombination yield and its associated gradients. In the Appendix we evaluate the number of required matrix multiplications (scaling with $N^2$), leading to the key insight that for the considered scenario and typical $N$s and problem sizes the cost of gradient/expectation value evaluation will exceed that of evaluating the elementary propagators and their derivatives (scaling with $N$). To allow for efficiently optimizing the reaction yield of radical pairs, we thus suggest the adaptation of the gradient-based optimization procedure as illustrated in Fig.~\ref{fig:sketch}. 

First, we argue that the optimization can be realized in practice in terms of $B$ disjoint blocks of $N/B$ time-steps each (time blocking). Second, we suggest that a sparsely sampled, i.e.\ crude, approximation of the reaction yield based on $N/(BS)$ samples per block, i.e.\ sampling every $S$-th time-step, is sufficient to adequately optimize the reaction yield. Here, $B$ and $S$ are integers that divide $N$ and $N/B$, respectively. While this approach formally does not yield the optimum of the yield as defined in eq.~(\ref{eq:yield}), henceforth referred to as the complete optimum, we argue that, for suitably chosen $B$ and $S$, yields optimized through sparsely sampled time blocking, realized at a fraction of computational effort, are sufficiently close to the true optimum. Third, to efficiently calculate the exponential propagators and the associated Fr\'echet derivatives, we use a scaling and squaring approach in combination with a Trotter-Suzuki splitting of the elementary propagator, as detailed in the Appendix. Lastly, to accelerate the optimization we utilise L-BFGS-B, a limited memory variant of the Broyden–Fletcher–Goldfarb–Shanno algorithm, which preconditions gradients with curvature information in lieu of steepest descent \cite{qLbfgs}.

\subsection{Protocol applied to a prototypical spin system} 

\begin{figure} [b]
	\centering
	\includegraphics[width=0.75\columnwidth]{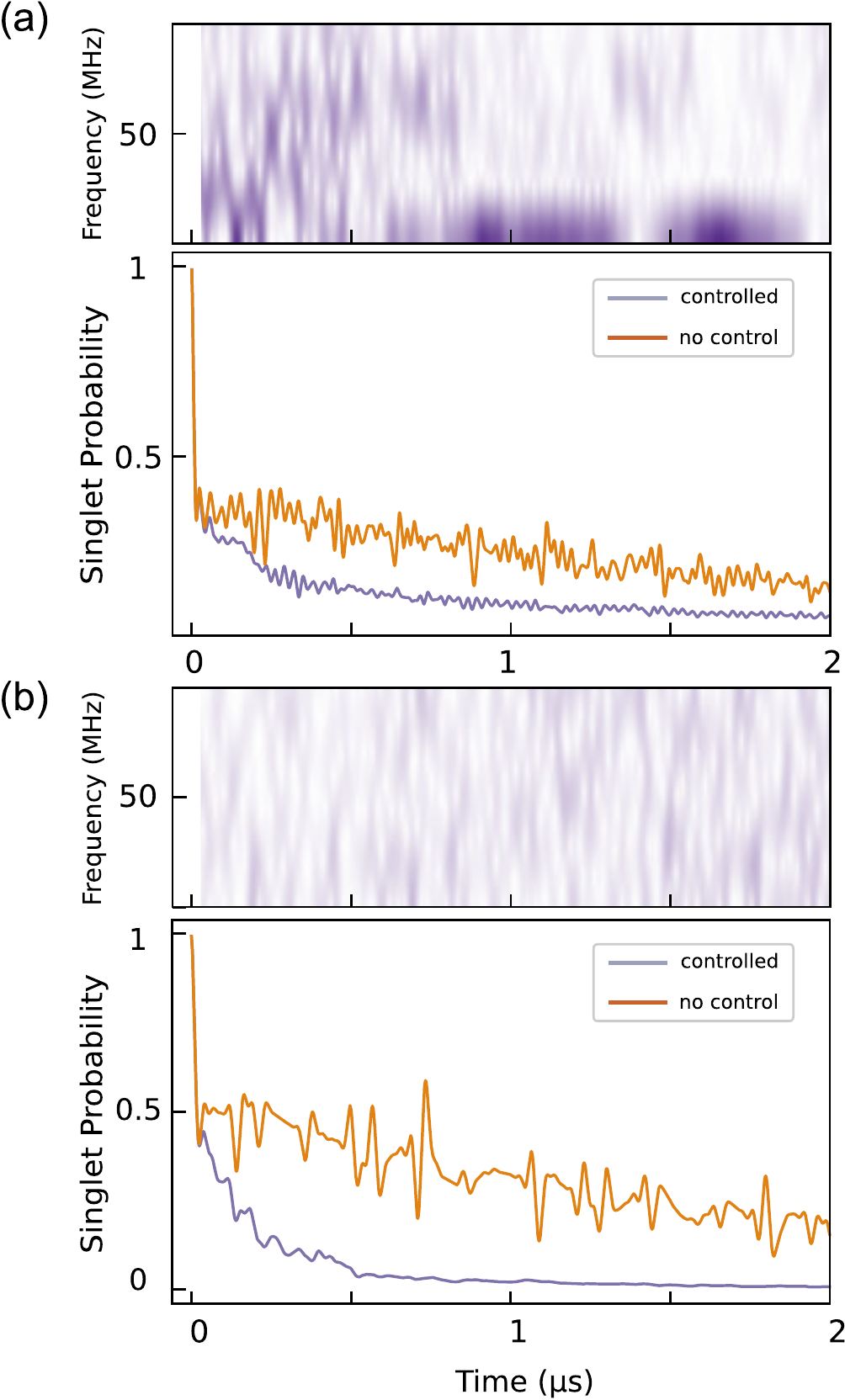}
	\caption{Controlled recombination dynamics of a 7-spin radical pair system under $0.5\mskip3mu$mT and $10\mskip3mu$mT static biasing fields in (a) and (b), respectively. The plots depict the survived singlet probability (cf. eq.~\ref{eq:gradients}) with (blue) and without (orange) coherent control, scaled to remove the spin-independent decay, i.e.\ \( p_{\rm S}(t)\exp(k_f t) \). The recombination yield was minimized over \( N = 5000 \) time-steps split into \( B = 2 \) blocks of \( 2500 \) time-steps each, sampled sparsely at \( S=5 \) and \( S=10 \) for (a) and (b), respectively. Accompanying spectrograms display the control field magnitudes, computed via discrete Fourier transforms of 64 points, overlapped by 60 points, and windowed using a Tukey function with a shape parameter of 0.25. Using a control amplitude of $0.25\mskip3mu$mT, the singlet recombination yield could be reduced to $0.156$ (from $0.269$ without control) and $0.097$ (from $0.346$) for the two biasing fields.}
	\label{fig:maeda}
\vspace{-0.5em}
\end{figure}

We consider the minimization of singlet recombination yield for a generalization of the prototypical radical pair studied earlier in the work by Masuzawa et al.\ \cite{mae}, where coherent control of a fidelity measure, rather than the reaction yield, was achieved in the simpler high-field scenario using time-local optimization. Specifically, we assume a radical pair comprising 7 spin-$\frac{1}{2}$ particles, coupled through isotropic hyperfine ($0.2$, $0.5$, $1\mskip3mu$mT and $0.2$ \& $0.3\mskip3mu$mT for the two radicals, respectively) and exchange interactions ($j_{ex}/(2\pi) = 1\mskip3mu$MHz), undergoing spin-selective recombination in the singlet state with rate constant $k_b = 1 \mskip3mu\mu$s$^{-1}$ or spin independent escape with rate constant $k_f = 1 \mskip3mu\mu$s$^{-1}$, such that $k_{\rm S} = k_b + k_f$ and $k_{\rm T} = k_f$. Given the high computational demands of completely optimizing the 7-spin system, we have additionally studied the 5-spin system resulting from the 7-spin system by retaining the three largest hyperfine coupling constants. Unlike Masuzawa et al.\ \cite{mae}, we seek to directly minimise the singlet recombination yield
\begin{align}
Y_b = \frac{k_b}{k_b + k_f} Y_{\rm S}
\end{align}
through a piece-wise constant control field applied perpendicular to a static biasing field, i.e.\ $L = 1$ and $V\equiv V_1 = \omega_1 (\hat{ S}_{1,x} + \hat{S}_{2,X})$, where $\omega_1$ is the maximal Rabi frequency of the control field; control amplitudes are subject to $|{u}_{1,n}|\equiv|{u}_n| \leq 1$. We chose a discretization time step of $1\mskip3mu$ns and controlled the first $5\mu\mskip3mu$s ($N = 5000$) after initialization of the spin system in the singlet electronic state (thus accounting for more than 99.3\% of radical decay; final yields still evaluated in the $t\to\infty \sim t_{max}$ limit with $t_{max}=14\mu\mskip3mu$s). Fig.~\ref{fig:maeda} illustrates exemplary time dependent singlet probabilities, controlled and unperturbed, associated with this scenario for the biasing field intensities $B_0 =0.5\mskip3mu$mT and $10\mskip3mu$mT. 

\begin{figure*}
	\centering
	\includegraphics[width=1.65\columnwidth]{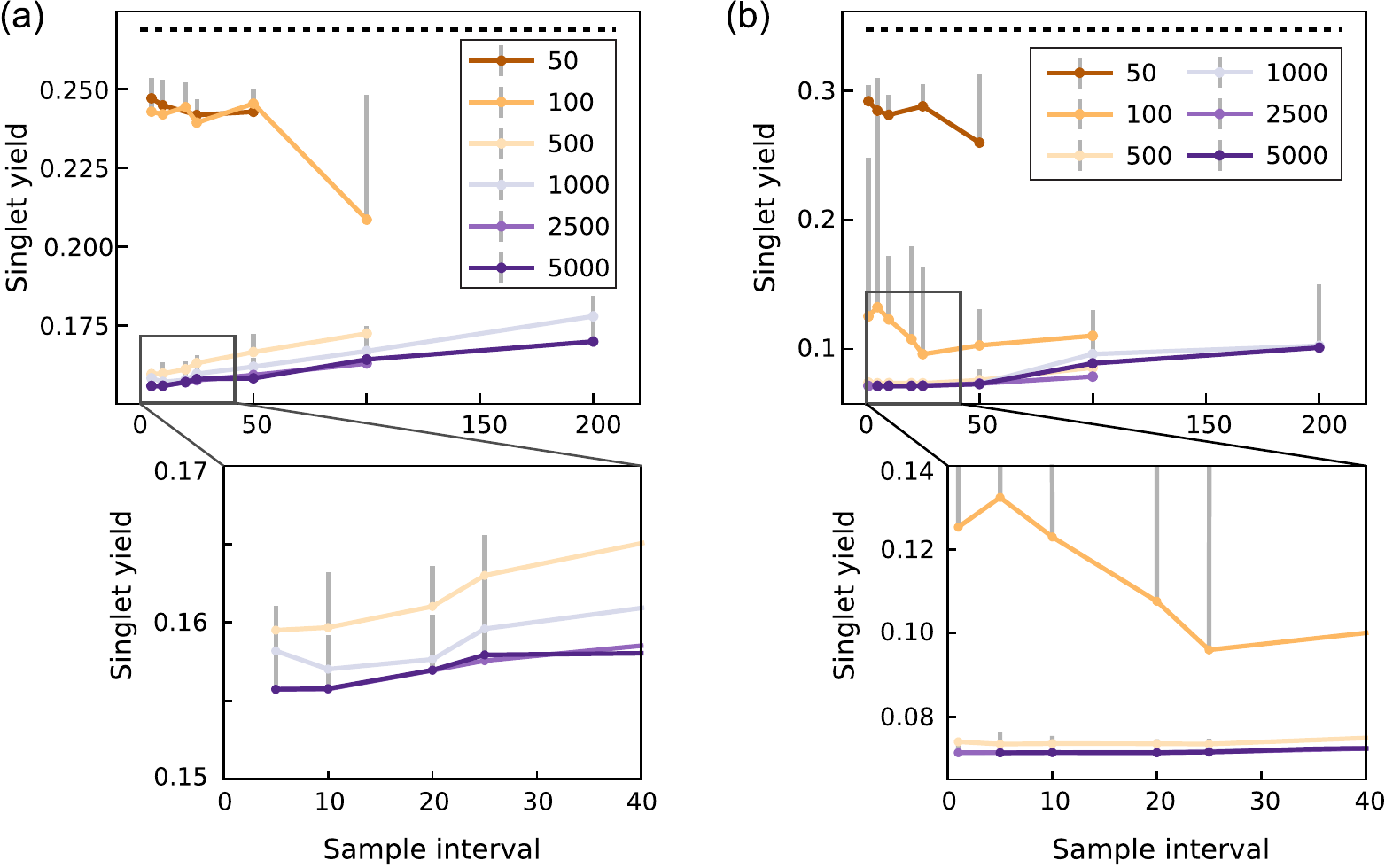}		
	\caption{For the 7 and 5-spin model systems, a maximal control field amplitude of $0.25\mskip3mu$mT, with biasing fields of $0.5$ and $10\mskip3mu$mT for (a) and (b), respectively, we show singlet recombination yield as a function of $S$ for yield minimizations using $B$ blocks of $N/B$ steps each, the latter value reported in the captions ($N=5000$). The plots present the best minimization result from $6$ independent replications with randomly chosen initial control amplitudes (chosen from a normal random distribution truncated at $\pm 1$); with dots indicate the 75-percentile range. The best optimization results allowed a reduction of the singlet recombination yield from $0.269$ to $0.156$ and from $0.348$ to $0.071$ for (a) and (b), respectively.}
 	\label{fig:maeda2}
\vspace{-1em}
\end{figure*} 

Efficient evaluation of single time-step propagators, and their Fr\'echet derivatives, was made possible via iterated Trotter-Suzuki splittings, and time blocking and sparse sampling of reaction yields, as introduced above. In view of Fig.~\ref{fig:maeda2} and comparable datasets for other biasing fields provided in the Supplementary Material (cf.\ Fig.~S4), it is apparent that excellent control results can be achieved for all $N/B$ block sizes exceeding $500$. Neither $B = 1$ and $N = B$ are strict necessities for practical control of the reaction outcome, substantially smaller blocks yield comparable optimization results at a fractional cost. On the other hand, block sizes of $50$ and $100$ are unsatisfactory, indicating that a time-local optimization is insufficient to yield optimal control results in the low to intermediate field region. For these inadequate block sizes, the results of the utilised L-BFGS-B optimization is furthermore strongly dependent on the initial condition. Note that intermediate block sizes $N/B$ are occasionally preferred over larger optimization blocks, as the latter are less prone to get stuck in local minima, such that the overall optimization outcome can be superior. This is, for example, seen in Fig.~S4 for a biasing field of $1\mskip3mu$mT, for which the best result of 6 replicated optimizations with $B=2$ and $B=5$ happened to surpass those with $B=1$. Note furthermore that the optimization results for smaller biasing fields, e.g.\ $0.5\mskip3mu$mT, is more sensitive to the sampling interval, $S$, than for $10\mskip3mu$mT. This is not surprising in view of the more varied singlet probability as a function of time observed for the lower field. This is encouraging insofar as more complex spin systems with diverse hyperfine interactions are expected to be less oscillatory. In any case, it appears that sampling intervals $S$ of $25$ and less are adequate here. 

For the 7-spin system we found it impractical to realize a complete set of optimizations for $S=1$ and $B=1$, i.e.\ complete, unblocked optimization, as optimization runs initiated from random configurations required more than a week to finish on the hardware utilized. However, a complete optimization starting from the best blocked optimization finished swiftly for the used tolerances without yielding a significant additional improvement. Furthermore, a single complete optimization yielded results comparable to those of the good blocked and sparsely sampled optimizations. For the smaller 5-spin system system, it is apparent that the sparsely sampled, blocked optimizations smoothly converge to the completely sampled optimum and that, again, $B > 1$ and $S > 1$ allow excellent practical optimization outcomes relative to this (expensive) limit. See Fig.~\ref{fig:maeda2}b and the Supplemental Material for these and further optimization results achieved for the 5 and 7-spin systems in weak magnetic fields, including the geomagnetic field (cf.\ Fig.~S1 and Fig.~S2). 

\begin{figure}
	\centering
	\includegraphics[width=0.8\columnwidth]{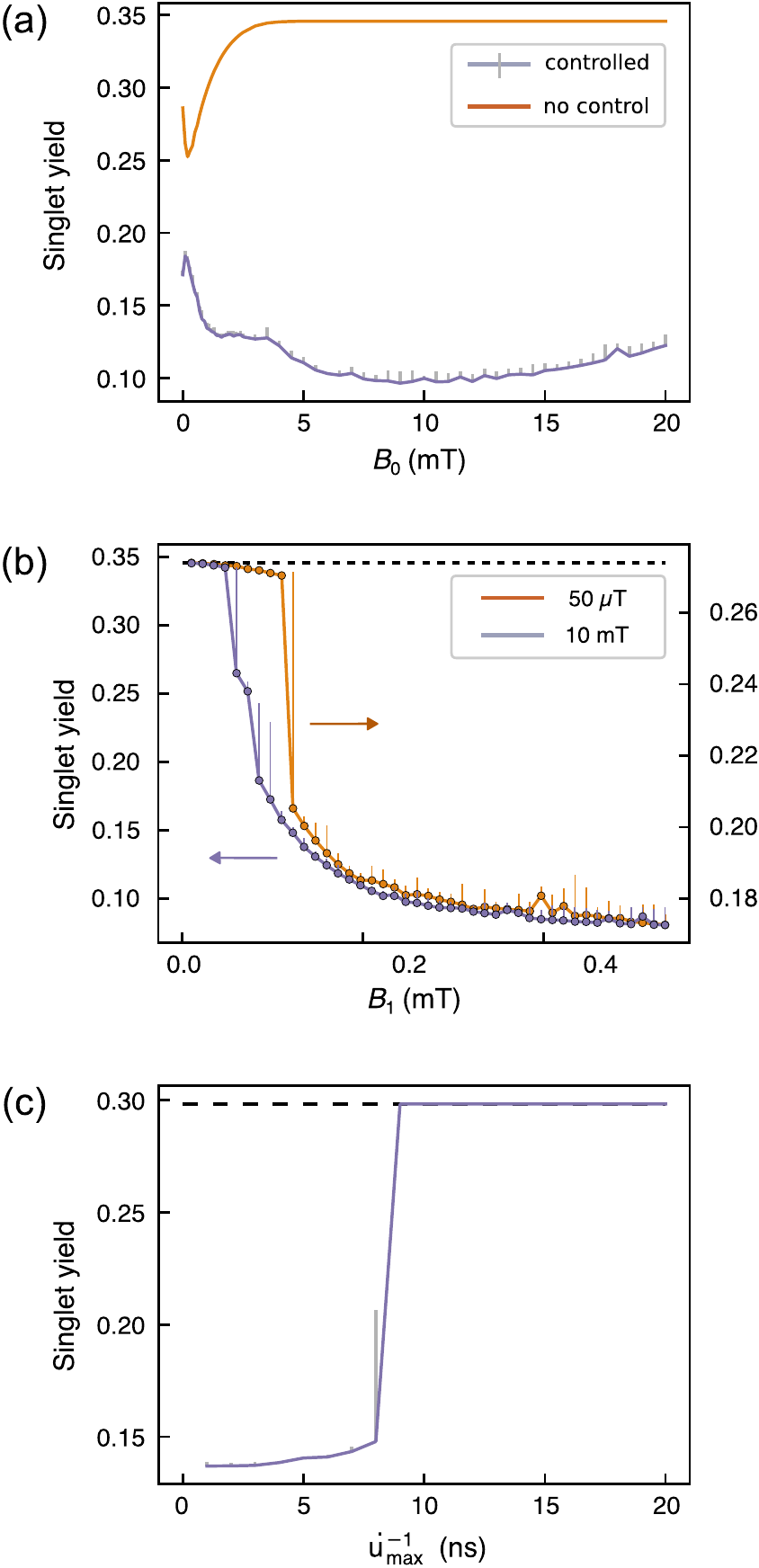}		
	\caption{Dependence on the optimization outcome on the field strength of the biasing field and driving field, and the control speed for the 7-spin model system. In (a), we consider a $0.25\mskip3mu$mT control field, and show yield minimization against increasing $B_{0}$ (blue) in comparison to the un-controlled reference (orange). In (b) we show yield minimization against increasing control field amplitude, $B_{1}$, for a biasing field of $B_0 = 10\mskip3mu$mT and $50\mskip3mu\mu$T. In (c) we show the dependence of the singlet recombination yield at $B_0 = 1\mskip3mu$mT on the speed limit $\dot{u}_{max}^{-1}$. For all optimizations, $N=5000$, $B=5$ and $S=25$.}
 	\label{fig:maeda3}
\vspace{-2em}
\end{figure} 

With the approach established, we set out to explore the prospects offered by controlling radical pair reaction yields in the weak to intermediate field case. Fig.~\ref{fig:maeda3}a illustrates the results of minimizing the singlet recombination yield as a function of the applied static magnetic field relative to the unperturbed scenario. These results were realized using $5$ blocks of $1000$ time steps of $1\mskip3mu$ns and sampling the recombination yield from every $25$th time point. The maximal control amplitude was $0.25\mskip3mu$mT. The plotted whiskers indicate 75-percentiles from 7 optimizations starting from random initial controls. Evidently, the recombination yield can be markedly reduced for all applied fields, the lowest yield realizable being $0.0966$, realized for a biasing field of $9\mskip3mu$mT. This amounts to a reduction by more than a factor of $3.5$ compared to the uncontrolled yield at the same field ($0.346$). Generally, it is remarkable that the yields controlled for minimal recombination are all markedly lower than the minimal recombination yield of $0.252$ realizable by applying a static magnetic field of $0.2\mskip3mu$mT, i.e.\ the field matching the dip in the singlet yield attributed to the low-field effect.

We have further analysed the dependence of the optimization outcome on the strength of the control field. Fig.~\ref{fig:maeda3}b shows $Y_b$ as a function of the maximal control amplitude for the case of a static biasing field of $10\mskip3mu$mT and $50\mskip3mu\mu$T, again realized by optimizing $5$ blocks of $1000$ time steps, sampling every $25^{\mathrm{th}}$. Appreciable minimization for $B_0=10\mskip3mu$mT requires control amplitudes of $0.1\mskip3mu$mT; controls with field strength below $\approx 50\mskip3mu\mu$T do not elicit appreciable effects. This finding is in line with the well-studied effects of radio-frequency magnetic fields on radical pairs. In this scenario too, a resonant perturbation can only appreciably impact the spin evolution during the radical pair lifetime, $\tau$, if -$\gamma B_1/(2\pi)\tau \gtrapprox 1$, i.e.\ if the lifetime is sufficiently large to permit at least one Larmor precession in the field associated with the perturbation. As here $k_f^{-1} = 1\mskip3mu\mu$s and $k_b^{-1} = 1\mskip3mu\mu$s, we can expect effects for $B_1$ exceeding 36$\mskip3mu\mu$T to 72$\mskip3mu\mu$T to become effective, which is in tentative agreement with the observed $B_1$-dependence. On the other hand, driving fields larger than $0.1\mskip3mu$mT allow very significant reductions of the recombination yield. It is interesting to note that for controlling reaction outcomes in the geomagnetic field ($B_0 = 50\mskip3mu\mu$T) slightly larger control fields exceeding $0.12\mskip3mu$mT are required.

Above we have assumed that controls are applied with a time-resolution of $\Delta t= 1\mskip3mu$ns. While AWGs are available with sufficient sample rates and analog bandwidths  (e.g.\ $50\mskip3mu$GS/s at $15\mskip3mu$GHz bandwidth) and the delivery of broadband magnetic fields at the required intensity has been realized (e.g.\ for high-field ENDOR spectroscopy \cite{endor}), the question of the minimal bandwidth to control radical pair reactions is a pertinent one. To this end, we have studied the dependence of the control result on the ``speed limit” enforced on the control signal. Specifically, we stipulated that the controls must obey $|{u}_n-{u}_{n-1}|< \dot{u}_{max}\Delta t$.  For $B_0=1\mskip3mu$mT, Fig.~\ref{fig:maeda3}c summarizes the dependence of the singlet yield on this speed limit. We find that transition times of $\dot{u}_{max}^{-1} \lessapprox 8\mskip3mu$ns are required to elicit significant control; fast controls only deliver marginal additional gains. The minimal speed limit corresponds to a bandwidth of approximately $\frac{\dot{u}_{max}}{2} = 63\mskip3mu$MHz, which is of the order of, but smaller than, the magnitude of the spread of energy eigenvalues of the radical pair ($93\mskip3mu$MHz).  

\subsection{Protocol applied to a realistic exciplex-forming radical pair}

\begin{figure}
	\centering
	\includegraphics[width=0.8\columnwidth]{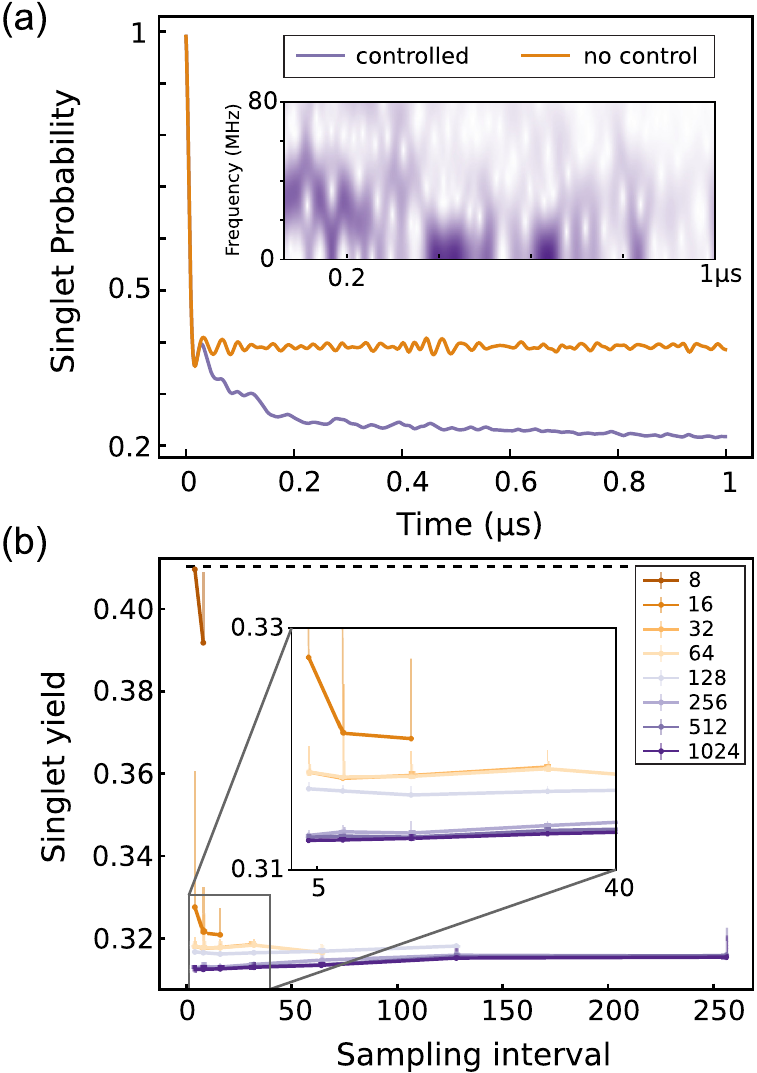}		
	\caption{Recombination yield minimization of the 18 spin PY/DCB system in a bias field of $1\mskip3mu$mT. (a) shows the singlet probability, scaled to remove the spin-independent decay due to radical recombination, i.e.\ $p_{\rm S}(t)\exp(kt)$, with and without control. Control amplitudes were optimized using $32$ samples and $1024$ time-steps of $1\mskip3mu$ns; a spectrogram of the control field magnitude is shown as insert. (b) shows the final yield as realized using L-BFGS-B complemented gradient optimization for increasing sampling intervals and various block sizes, $N/B$, as indicated in the legend.}
	\label{fig:pyDCB}
\vspace{-1em}
\end{figure}

We now present results for coherently controlling the symmetric recombination of a large spin system consisting of 18 spins. We consider the pyrene/para-dicyanobenzene (PY/DCB) exciplex-forming donor-acceptor systems in a solvent of moderate polarity \cite{kt11, hoang14, exci}. The magnetosensitive, delayed fluorescence emission of this system has been widely studied. To make this large scale simulation tractable, we limit ourselves to the scenario of homogeneous decay, i.e.\ $k_{\rm S}= k_{\rm T}=k$, and neglect inter-radical interactions. While this may appear crude at first glance, models typical of this have seen considerable success in the interpretation of experimental data \cite{hayashi04}. The approximations are adequate since for the majority of its lifetime, the radical pair is in fact diffusively well separated, and the moderate reaction asymmetry of the real system (resulting from different electron transfer rate constants in the singlet and triplet state; the latter to the locally excited triplet state) has all but a small effect on the magnetic field sensitivity. Thus, in this limit, the singlet probability can be expressed as 
\begin{align}
p_{\rm S}(t) = e^{-kt} \left(\frac{1}{4}+\sum_{\alpha\beta\in\{x,y,z\}} R^{(1)}_{\alpha,\beta}(t)R^{(2)}_{\alpha\beta}(t)\right) ,
\end{align}
where
\begin{align}
R^{(i)}_{\alpha\beta}(t) = \frac{1}{Z_i}{\rm Tr}\left[\hat{S}_{i\alpha}(0)\hat{S}_{i\beta}(t)\right] \nonumber \\ 
= \frac{1}{Z_i}{\rm Tr}\left[\hat{S}_{i\alpha}\hat{U}^{\dagger}(t,0) \hat{S}_{i\beta}\hat{U}(t,0) \right]
\end{align}
are the electron spin correlation tensors for each spin in radical $i$, which are defined or calculable in the Hilbert space of the individual radicals. Here, $\hat{U}$ accounts for the coherent evolution only, i.e.\ $\hat{A}(T) = \hat{H}(T)$ (cf.\  eq.~(\ref{eq:unitary})). Thus, control of the radical pair reaction yield can be realized by separately calculating the gradients for each of the spin correlation tensors of two recombining radicals, and assembling the gradient of the reaction yield using the chain rule. Again, we follow the ideas of blocking the time axis and sparse sampling to increase the efficiency. In addition to the Trotter-Suzuki splitting discussed above, we use the fact that the Hamiltonian is only a function of the total nuclear spin associated with groups of completely equivalent spins, which permits an efficient calculation in the ``coupled representation'' \cite{atkins19}. Fig.~\ref{fig:pyDCB} shows the minimization of the PY/DCB singlet recombination yield  for $k^{-1} = 200\mskip3mu$ns, a control amplitude of $0.2\mskip3mu$mT, and a static basing field of $10\mskip3mu$mT. The control was extended to the first $1024$ time steps of $1\mskip3mu$ns each. Varying block and sample sizes, again confirms practically sufficient fidelity from non-complete optimization, as is demonstrated in Fig.~\ref{fig:pyDCB}b (also see Fig.~S3). In our view, exciplex-systems of the type discussed here would be ideally suited to realize first proof-of-principle experimental realizations of OQC of radical pair reaction yields \cite{exci, kt11, hoang14, Rugg2019, Mims2021} by building on the time-resolved MFE measurements and studies of radio-frequency effects that have already been realized.

\subsection{Applications to Quantum Biology} 

Many observations of biochemical magnetosensitivity appear to be attributed to radical pair intermediates in complex biophysical environments \cite{hore16, simon2022}. We anticipate that the control of radical pair reactions will develop into a versatile applied field, particularly in the spin-biological context. Static and oscillatory magnetic fields can and have been used to perturb such systems, but these na\"ive controls are neither selective for specific pathways nor can reaction yields be changed drastically. But our OQC approach promises distinguishability of reaction pathways involving different radical pairs and enlarged effect sizes in weak biasing fields, such as the geomagnetic field. We demonstrate this potential here in terms of two selected examples with a quantum biology underpinning. 

\begin{figure}
	\centering
	\includegraphics[width=0.8\columnwidth]{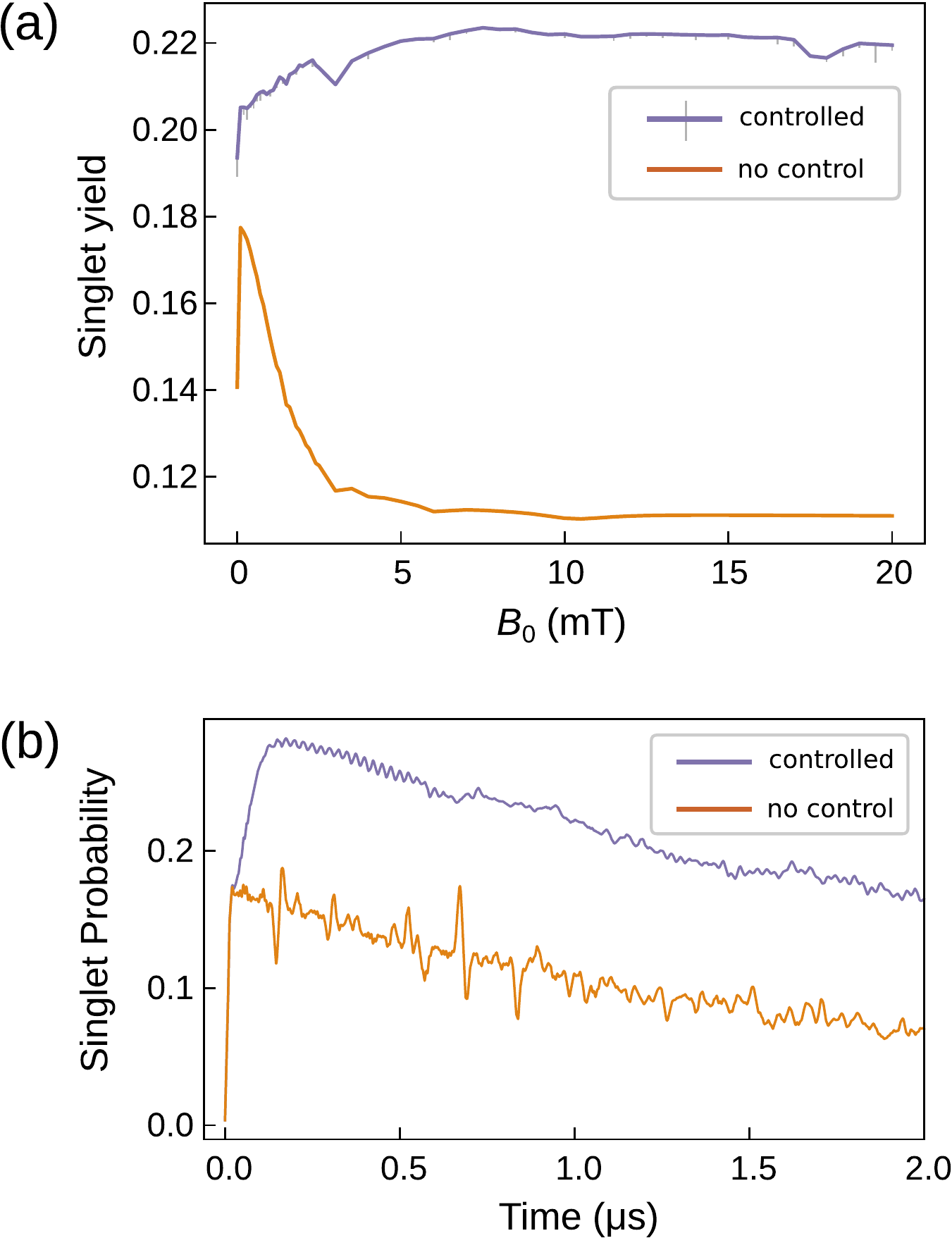}		
	\caption{Controlled and unperturbed singlet recombination yield of a flavin semiquinone/superoxide radical pair, abstracted as FADH$^{\bullet}$/Z$^{\bullet}$, initialized in the triplet state, as a function of the applied static biasing field. The radical pair was assumed to undergo a spin-independent decay/escape reaction with rate constant $k_f = 1\mskip3mu\mu$s$^{-1}$ and spin-selective recombination in the singlet state with rate constant $k_b = 1\mskip3mu\mu$s$^{-1}$. The first $5\mskip3mu\mu$s after radical pair generation were controlled using a step-wise constant control field with $1\mskip3mu$ns time resolution and a maximal amplitude $0.25\mskip3mu$mT, applied perpendicular to the static magnetic field. The control was optimized using the L-BFGS-B algorithm applied to $5$ blocks of $1\mskip3mu\mu$s, sampling the yield every $25\mskip3mu$ns. Optimizations were repeated $5$ to $6$ times starting from random initial controls; with the error bars indicating the $75$-percentile range of best results (the variability between runs is noticeably small compared to the change in recombination yield produced through control). In b), we show the time-dependence of the singlet probability of the radical pair in a $4\mskip3mu$mT biasing field in absence and presence of the optimized control field, scaled to remove the spin-independent decay, i.e.\ \( p_{\rm S}(t)\exp(k_f t) \). The associated singlet recombination yields amount to $0.218$ and $0.115$ with and without controls, respectively.}
	\label{fig:avian}
\vspace{-1em}
\end{figure}

The flavin semiquinone (FADH$^{\bullet}$)/superoxide (O$_2^{\bullet-}$) radical pair has been suggested to underpin various biomagnetic processes, as implicated by MFEs in neurogenesis \cite{Ramsay2022} and cellular energetics \cite{Usselman2016}. Being of so-called reference-probe type, this radical pair is renowned for its sensitivity to weak magnetic fields \cite{Lee2014}, but questions concerning its potentially fast spin relaxation remain \cite{hogben2009}. We consider the limit of slow spin relaxation, conventionally referred to as the FADH$^{\bullet}$/Z$^{\bullet}$-model. Here, Z$^{\bullet}$ denotes a radical that is devoid of hyperfine interactions and slowly relaxing, i.e.\ an idealization of O$_2^{\bullet-}$, as it might be approached in the real system through immobilization. Fig.~\ref{fig:avian} shows the ability of our approach to maximize the recombination yield of the triplet-born radical pair. This model, with parameters taken from \cite{Deviers2022}, exhibits a large low-field effect in response to static magnetic field (orange line). Controlling the recombination (blue line) allows to boost the recombination efficiency, realizing recombination yields that are not achievable by the application of static magnetic fields of any strength. In a biophysical context, this would lead to a decrease of superoxide release from flavin re-oxidation reactions, thereby reducing the concentration of free superoxide. Superoxide is a central reactive oxygen species that assumes important roles as a signalling molecule, thus linking radical pair reactions to central biological control pathways \cite{Winterbourn2020}. Since magnetosensitivity in a physiological setting is currently not well understood beyond circumstantial evidence, the latter is a more distant aim, albeit an auspicious one, because it could allow targeting specific redox pathways in isolation for various applications, e.g. wound healing \cite{cellMFE} or the suppression of the adverse effects of hypomagnetic field \cite{Zhang2021, Ramsay2022} exposure for space travel. Other future applications might be the control of chemical reactions for synthetic purposes, e.g. in the context of flavin photocatalysis \cite{flavC}.

In sensory biology, a certain consensus has been reached attributing an inclination compass in birds and various other species to the proposed radical pair mechanism involving the flavo-enzyme cryptochrome \cite{hore16}. However, the identity of the actual magnetosensitive radical pair is still subject to debate. \emph{In vitro} experiments on cryptochromes have demonstrated magnetosensitivity of a photo-induced radical pair comprising the flavin anion radical (FAD$^{\bullet-}$) and a surface exposed tryptophan cation radical (W$^{\bullet+}$), implicating FAD$^{\bullet-}$/W$^{\bullet+}$ as the sensory species \cite{Xu2012}. However, magnetoreception appears to also be possible in the dark, suggesting the potential involvement of a re-oxidation pathway, implicating once more the flavin semiquinone/superoxide radical pair (FADH$^{\bullet}$/O$_2^{\bullet-}$) in magnetoreception. Furthermore, various alternative radical pairs have been suggested in theoretical studies demonstrating superior magnetic field sensitivity, e.g.\ for ascorbyl and Z$^{\bullet}$ containing radical pairs \cite{Lee2014}. Given the multitude of suggestions and hypothesis, the distinguishability of various radical pairs in spin biology is a timely question. This question cannot be resolved by direct spectroscopic means as the radical pair is a transient intermediate in a complex biophysical environment in a living organism. We here demonstrate that OQC could in principle provide the sought insights. 

\begin{figure*}
	\centering
	\includegraphics[width=1.65\columnwidth]{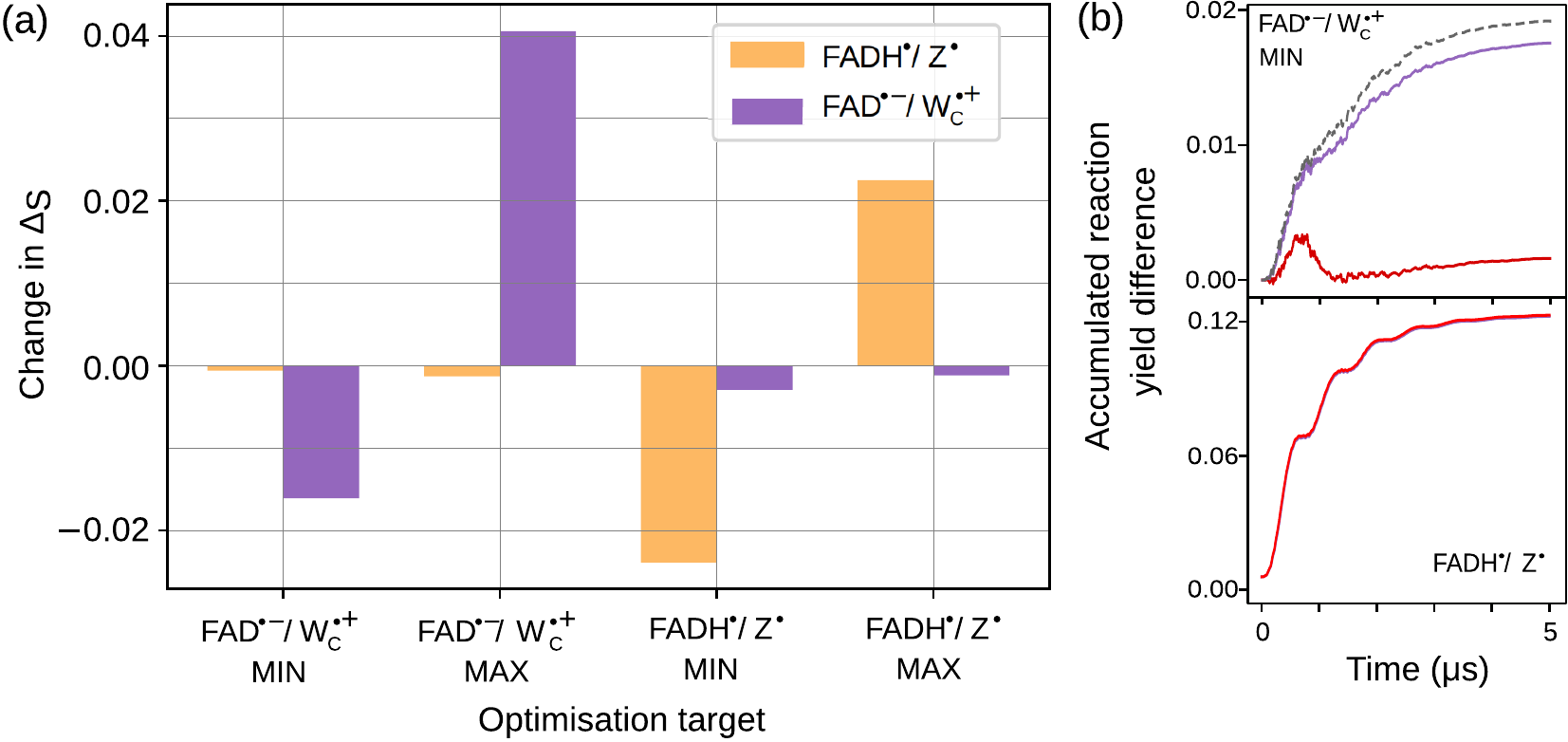}		
	\caption{Change in singlet yield contrast $\Delta_S$ for  FAD$^{\bullet-}$/W$^{\bullet+}$ and FADH$^{\bullet}$/Z$^{\bullet}$ radical pairs subject to controls optimized to maximise or minimise the singlet recombination yield contrast $\Delta_S$ in the geomagnetic field for one of the pairs (as indicated in the axis labels). The geomagnetic field was assumed as $50\mskip3mu\mu$T and the radical pair lifetime $1\mskip3mu\mu$s, i.e.\ $k_S = k_T = 1\mskip3mu\mu$s$^{-1}$. The control was applied parallel to the static field with an amplitude of $12.5\mskip3mu\mu$T. The first $5\mskip3mu\mu$s after radical pair generation in the singlet state were controlled with a time-step of $1\mskip3mu$ns. Yields were calculated by integrating to $t_{\text{max}} = 14\mskip3mu\mu$s. The optimization was realized through $40$ steps of steepest descent/ascent. The unperturbed singlet yield contrasts are $\Delta_S = 0.017$ and $0.13$ for  FAD$^{\bullet-}$/W$^{\bullet+}$ and FADH$^{\bullet}$/Z$^{\bullet}$, respectively. In b), we show the accumulated reaction yield differences for each, $k_S \int_0^t p_S(n_{\text{max}},t) - p_S(n_{\text{min}},t) \, dt$, i.e.\ the accumulated yield difference up to time $t$ as a function of time for the  FAD$^{\bullet-}$/W$^{\bullet+}$ (top panel) and FADH$^{\bullet}$/Z$^{\bullet}$ (bottom panel) radical pair. The red lines for each radical pair correspond to the case of the radical pair dynamics being perturbed by the OQC-derived perturbation targeting the minimization of $\Delta_S$ for  FAD$^{\bullet-}$/W$^{\bullet+}$. The FADH$^{\bullet}$/Z$^{\bullet}$ pair is practically unresponsive to this perturbation (bottom panel), while FAD$^{\bullet-}$/W$^{\bullet+}$’s directional sensitivity is mostly attenuated. For FAD$^{\bullet-}$/W$^{\bullet+}$, the grey line applies to the scenario of applying a harmonic perturbation with frequency $1.4\mskip3mu$MHz, demonstrating that designed perturbations are significantly more disruptive than unspecific perturbations.}
	\label{fig:apply}
\vspace{-1em}
\end{figure*}

We considered models of the FAD$^{\bullet-}$/W$^{\bullet+}$ and FADH$^{\bullet}$/Z$^{\bullet}$ radical pairs, including the anisotropic hyperfine interactions of the dominant nuclear spins in a relative orientation as found in a cryptochrome (parameters taken from \cite{Babcock2021}). The recombination yield of this system depends on the orientation of the geomagnetic field relative to the protein frame. The difference of the maximal and minimal yield, realized for field directions ${\bf n}_{\text{max}}$ and ${\bf n}_{\text{min}}$, i.e.\ $\Delta_S = Y_S({\bf n}_{\text{max}}) - Y_S({\bf n}_{\text{min}})$ is considered a measure of the sensitivity of the radical pair-based compass. In Fig.~\ref{fig:apply}, we demonstrate the ability of OQC to control, i.e.\ maximize or minimize, this performance parameter for the two radical pairs considered. The control fields were applied parallel to the geomagnetic field. Excitation profiles were optimized for one radical pair and its performance subsequently evaluated in both. We find that OQC can significantly alter, enhance or attenuate, $\Delta_S$ for both radical pairs. Importantly, sequences optimized for one pair turn out to only have a minor effect of the spin dynamics of the other. Thus, the control fields are selective, permitting to distinguish the two models insofar as they induce a large effect on the singlet recombination yield contrast in the targeted pair and a negligible effect in the competitor. In particular, the controls derived for FAD$^{\bullet-}$/W$^{\bullet+}$ appear as an apt tool to realize such an experiment, because the contrast $\Delta_S$ is strongly altered, the system specificity high, the time of radical pair generation is controllable using pulsed photo-excitation, and the parameters pertaining to FAD$^{\bullet-}$/W$^{\bullet+}$ are best known, as they are available from \emph{in vitro} experiments and derivable from crystal structures and computational chemistry methods. Clearly, here we limit ourselves to only demonstrate principal feasibility; more detailed models will have to be studied before instigating actual experiments.

The robustness of our derived control pulses to small changes is central for future applications. While control engineering generally aims to capture the intricacies of the control target as closely as possible, in practice our knowledge of the actual spin Hamiltonian parameters \cite{optPar} might be limited, or the available computational resources might restrict the optimization to a simplified spin system comprising only the strongest hyperfine-coupled nuclear spins. In this context, controls derived from simplified systems, in particular those not accounting for weakly coupled nuclear spins or spin relaxation processes, are crucial. As evidenced in the Supplemental Material, controls derived using our approach are indeed robust when applied to simulations accounting for the relevant inhomogeneous hyperfine interactions via the Schulten-Wolynes framework \cite{sch_wol} (cf.\ Fig.~S5). The controls are also found to be robust against noise in the form of random field fluctuations, which induce decoherence and spin relaxation (cf.\ Fig.\ S6). We particularly find that controls derived for idealized systems retain their unique property of permitting recombination yields that cannot be realized using static magnetic field of any flux density even in the presence of spin relaxation and unresolved hyperfine interactions.

\section{Discussion}

In electronic and nuclear spins of organic radical pairs, coherence lifetimes on the order of $100\mskip3mu$ns to microseconds are typical, even in complex biophysical environments \cite{kt16}, opening a window of opportunity to realize control by applying time-dependent magnetic fields to manipulate reaction outcomes. Yet, the approach is still in its infancy. Experimental approaches have been limited to perturbing radical pair spin systems by radio-frequency magnetic fields \cite{mani22, engels14}. While positioned far from optimal control, these experiments are crucial as they demonstrate the principal possibility of influencing radical pair dynamics by RF-magnetic fields in weak magnetic fields. Examples of RF-magnetic field effects are furthermore well-established for spin-biological radical pairs \emph{in vivo} \cite{mani22}, such as those supposedly underpinning the avian compass sense. The latter are particularly noteworthy, because surprisingly weak RF-magnetic fields have been found to interrupt magnetoreception, suggesting an exquisitely sensitive underlying system of remarkable coherence time. 

Optimal quantum control protocols could be used to drive chemical and spin-biological systems in a controlled and selective way, whereby reaction outcomes not attainable by static magnetic field perturbations alone become accessible. For quantum information processing in near term quantum hardware, among the many hurdles faced in counteracting environment-induced decoherence is the need to realize precise timing of quantum gate operations \cite{qGate23, impTC} in nanosecond timescales. To evade this bottleneck, precise or adaptive \cite{adQC, fmQC, cont22b} quantum control protocols may be harnessed to change the frequency of an applied control field dynamically throughout its pulse duration.  But this is often hard to realize in practice, since formulating the appropriate optimal quantum control protocols is often highly non-trivial. In nature, radical pairs seem to evade the effects of decoherence to retain exceptionally long coherence times, resulting in exquisite and unparalleled magnetometric sensitivity even in extremely noisy biological settings \cite{smith2024optimality}. Thus, optimal quantum control protocols, formulated with the primary aim of reproducing this sensitivity in spin-chemical reactions in solution, could also inform methods for similarly realizing artificial or bio-molecular \cite{qiQC, mani22} qubits highly sensitive to applied magnetic fields. This could translate to further applications \cite{brieg10, qGate23, bio-comp22a, bio-comp22b} in both quantum metrology and quantum information processing. In particular, potential engineering principles derived from gleaning deeper physical insights gained from artificially driving radical pair reactions could conceivably inspire a new generation of quantum enhanced sensing technologies for magnetometry boasting heightened sensitivity and robustness against noise \cite{brieg10, Aiello2013, vitalis2017, qcRp21, bio-comp22b, Smith2022, Winkler2023}. But to realize this goal, and future applications of optimal quantum control in a potentially biophysical context, efficient approaches for control design are vital.

The control of radical pair reactions is distinct from the majority of OQC approaches in the sense that the optimization target is not defined at a particular instant of time, but encoded in the time-evolution from radical-pair creation to its eventual decay, $t \in [0,\infty)$, i.e.\ the reaction yield. Here, we have generalized the GRAPE-paradigm of using piece-wise constant control amplitudes to realize reaction control. The suggested algorithm aims to overcome limitations of previous approaches by targeting actual reaction yields (rather than continuously updating the control to optimize a fidelity function involving observables that commute with the Hamiltonian or the singlet probability at one instance of time) and permitting control in weak magnetic fields. A time-global update scheme of control parameters (as opposed to the time-local approach from \cite{sugi03, mae, mae2}), central to GRAPE, has here been found essential to realize the latter efficiently. The computational cost of controlling radical pair reaction yields by time-global optimization techniques such as GRAPE has been viewed as prohibitive \cite{mae}. And indeed, significant computational costs are associated with the evaluation of the numerous single time-step propagators and their Fr\'echet derivatives, and, for the optimization of reaction yields in particular, the calculation of the time-integral recombination yield and its gradient through iterated matrix multiplications involving the former. We have introduced a block-optimization scheme and sparse sampling of the reaction yield to improve the efficiency of this step while retaining adequate control fidelity. To facilitate the efficient evaluation of propagators and their derivatives, we have utilized an iterated Trotter-Suzuki splitting. We further leverage curvature information of the loss function in lieu of standard steepest decent/ascent approaches to reduce the number of optimization steps required, as previously suggested \cite{qLbfgs}. 

The additional algorithmic modifications we incorporate into our GRAPE-inspired approach are vital for addressing the high computational costs, which had impeded earlier attempts to realize GRAPE-based reaction control of radical pair dynamics. We note that while sparse sampling of the reaction yield demonstrably provides a useful and computationally feasible proxy to the actual reaction yield, sampling of the singlet probability at a single moment in time is oftentimes misleading the optimization in weak magnetic fields as the increased/decreased probability at the chosen time does not necessarily correlate with increased/decreased reaction yield. On the other hand, as we have shown, sparse sampling at a moderate sampling density reduces the computational cost while yielding close to optimal optimization outcomes and a smooth progression for $B > 1$. We have further demonstrated the applicability of the approach as implemented here to realistically complex radical pair systems, such as the widely studied PY/DCB exciplex system subject to symmetric recombination. Our suggested approach can serve as a blueprint for future developments and optimizations. Although not explored here, further algorithmic modification could be attempted, such as bottom-up strategies for which the sampling density is increased as the optimization progresses. Additionally, the optimization blocks could be overlapped. Nonetheless, as our approach yielded control parameters that upon complete optimization did not change significantly subject to the utilized termination criteria of the optimizer, these additional improvements did not appear imminently pertinent. Although, they could further speed up the rate at which the target is reached. Additional computational efficiency could furthermore be realized by reformulating the singlet yield evaluation in terms of the propagation of wave-packets, in particular in combination with using an incomplete basis for the nuclear spin functions and trace sampling \cite{fay21}. We are currently developing such approaches with the aim of extending control to the complete open systems \cite{kosloff22} description of radical pair systems, including via the incorporation of direct control of the dissipation processes through incoherent control. As for the calculation of propagators and derivatives, higher order operator splitting approaches could be used and the order of the approach, i.e.\ accuracy, to the optimization process increased as the target function converges, as has been successfully demonstrated in \cite{cont22b} for a state transfer problem. However, since the generation of propagators and derivatives was not identified as the most time-consuming step for the considered problem sizes, which is related to the evaluation of reaction yields and gradients instead, here we have again refrained from such additional optimizations. From the reaction-control point of view, it is noteworthy that control has allowed us to suppress the singlet recombination to levels below those achievable by application of static magnetic fields of any strength. Furthermore, the achievable recombination yields are well below the $25\mskip3mu$\% yield that is expected to ensue for complete randomization of the spin states. This shows that the control fields here do not merely enhance spin relaxation, as one might expect, but drive the system to optimal reaction outcomes. 

Finally, we comment on the alternative approach of time-local optimization as introduced in \cite{mae}. Even with the described optimization tweaks, the presented GRAPE-based approach to radical pair reaction control remains computationally more demanding than the time-local approach. However, it captures the time-integrated nature of the reaction yield, the observable quantity, rather than merely the singlet probability at one moment in time, which is not necessarily a well-defined proxy of the overall recombination yield. Another limitation of the time-local approach is conceptual and related to the property of permitting only controls that at every moment of time increase the optimization target function. This property permits the straight-forward derivation of the optimal control amplitudes but constrains the optimizations by excluding controls that would permit a better overall optimization at the cost of a momentary reduction in optimality. We argue that this constraint is critical in the optimization of weak biasing fields. In the Supporting Material we provide a generalized version of the time-local algorithm from \cite{mae} that permits the optimization of the singlet probability, still for one instance in time, in the presence of spin selective recombination, as assumed here. Comparing this generalized algorithm to the GRAPE-based approach, we find that it is inferior for optimizing the recombination yield in weak biasing fields, generally requiring larger control amplitudes while still falling short in terms of optimization fidelities. Please refer to the Supporting Material for the detailed quantitative comparison undertaken for the 7-spin model system. As shown there, superiority of the GRAPE-based approach is realized not only when optimizing the singlet probability at a single time, but also when applying block-wise approach with successively redefined ``moving target''. While this time-blocking approach appears auspicious for the time-local approach too, the GRAPE-based approach delivers markedly better yield optimization in weak biasing fields.

The quantum control of radical pair reactions is still in its infancy. Our present work focuses on the control of the coherent evolution via Zeeman interactions, as this is a modality in principle addressable with currently available experimental tools without complex changes to the chemistry. We anticipate that optimal control of radical pair reactions will find applications in the context of spin-biological radical pair reactions, either as an analytical tool or to drive the systems towards desired outcomes. For example, for radical pair reactions implicated in magnetoreception, a key problem to be addressed concerns distinguishing between different incarnations of radical reactions to better identify the underlying mechanism in a complex biological environment \cite{kt19, qcRp21, nt21}. In this scenario, quantum control could allow us to derive control parameters that suppress magnetic field effects in one mechanism while leaving those of competing alternative models unaltered. Ultimately, controls could allow us to turn MFEs into a tool to potentially control reaction processes of physiological relevance for applications in biology and medicine, by facilitating selective stimulation or suppression of cellular functions linked to radical pair processes. We hope that this work, by building on established techniques from quantum control like GRAPE to go beyond existing local optimization techniques, will lay the groundwork for realizing this.

\section{Conclusions}

We have suggested and implemented a GRAPE-inspired approach to control reaction yields of radical pair recombination reactions through synchronized application of radio-frequency magnetic fields in a direction perpendicular to a static biasing field. We deviate from previous approaches by optimizing reaction yields, i.e.\ the actual experimental observables, instead of fidelity measures defined at a particular time instants. Computational efficiency has been realized by block optimization, sparse sampling, a split operator approach for calculating propagators and by utilizing curvature information in the non-linear optimization problem. The approach thus realized is applicable to radical pair systems of realistic complexity. We have demonstrated that control permits the reduction of the singlet recombination yield of a singlet-born model system to $10\mskip3mu$\%, a yield unattainable by static magnetic field application alone, when using moderate control amplitudes and bandwidth (cf.\ Fig.~\ref{fig:maeda3}).  We hope that our approach helps accelerate the adoption of control approaches for radical pair recombination reactions in low external fields. We anticipate that these future applications will utilize control beyond mere reaction control, for example to realize distinguishability of radical pairs in complex environments and for facilitating model selection for competing hypotheses based on experiments with optimized contrast power.

\begin{acknowledgments}
This work was supported by the Office of Naval Research (ONR Award Number N62909-21-1-2018), the Leverhulme Trust (RPG-2020-261), and the Engineering and Physical Sciences Research Council (EPSRC grants EP/V047175/1 and EP/X027376/1). We acknowledge use of University of Exeter’s HPC facility. 
\end{acknowledgments}
 
\setcounter{equation}{0}
\setcounter{figure}{0}
\renewcommand{\theequation}{A\arabic{equation}}
\renewcommand{\thefigure}{A\arabic{figure}}

\small

\section*{Appendix}
\subsection{Sparse Blocking}
The control of reaction yields via numerical approaches like GRAPE is expected to be costly. The computational demand is particularly large if, as we do here (and unlike \cite{mae, mae2}), the goal is to optimize recombination yields of radical pair reactions, which are time-integral quantities (cf. eq.~(\ref{eq:yield}); unlike the typical fidelity measures in quantum control applications pertaining to the implementation of a state transfer or a unitary operation for a fixed time $t$). The approach requires iteratively repeated matrix multiplications for evaluating the time-dependent singlet-probabilities that sum to the recombination yield (and its associated gradients). 

Formally, eq.~(\ref{eq:approx}) implies
\begin{align}
\sum_{n=1}^{N}[(n-1) +3] + nL[(n-1)+3]=O(N^3) 
\label{eq:mult}
\end{align}
matrix multiplications, i.e. for every $t_n$, $(n-1)$ matrix multiplications are required to evaluate $\hat{U}$, $(n-1)$ to evaluate $\hat{U}_{+}\hat{W}_{n,l}\hat{U}_{-}$ (via $L_{n,l}$) for every $l$ and $n$; and an additional three each to evaluate the expectation value/gradients. 
In practical implementations, the estimate in eq.~(\ref{eq:mult}) can be improved upon by preserving terms from previous $n$, i.e. $n-1$, allowing the update of $\hat{U}$ and $\hat{U}_{+}\hat{W}_{n,l}\hat{U}_{-}$ with a singular matrix multiplication (instead of $n-1$), thus requiring
\begin{align}
\sum_{n=1}^{N}4(1+ nL) = 2LN^2 +2(L+2)N
\end{align}
matrix multiplications. Nevertheless, due to the quadratic scaling of the evaluation of eq.~(\ref{eq:approxG}) with $N$, the gradient/expectation value evaluation will often exceed the cost of evaluating the elementary propagators and their derivatives (scaling as $N$) for large $N$ and thus be the limiting factor. 

To efficiently optimize the reaction yield of complex radical pairs, we suggest the adaptation of a gradient-based optimization procedure. We posit that the optimization can be realized in terms of $B$ disjoint blocks of $N/B$ time-steps each (time blocking), with the initial state taken as that of the system subject to the preceding optimized controls, sparsely sampled using $N/(BS)$ samples per block, as illustrated in Fig.~\ref{fig:sketch}. Here, $B$ and $S$ are integers that divide $N$ and $B$, respectively. The number of matrix multiplications in this approach is reduced to
\begin{align}
B\left(\sum_{n=1}^{N/B}(1+ nL)[1+3(\delta_{n \;\mathrm{mod}\; s,0})]\right) \nonumber \\
= B\left(\sum_{n=1}^{N/B}(1+ nL)+3\sum_{m=1}^{N/BS}(1+ LS)\right)\nonumber \\
= \frac{N^2L(S+3)}{2BS}+N\left(2L+1+\frac{3}{S}\right). 
\end{align}
The number of matrix multiplications necessary in general still scales quadratically with $N$, i.e.\ as $O(N{^2}/B)$. Linear scaling is only obtained in the limit that $B=N$, i.e.\ time-local optimization. But in practice, the block size can be chosen much smaller than $N$, i.e. $N/B \ll N$, to realize a substantial speedup, whilst delivering adequate practical optimization of the reaction yield. Furthermore, for $S\gg3$, the complexity of the algorithm approaches independence of $S$.

\subsection{Trotter-Suzuki Splitting}
To efficiently calculate the propagators and the associates Fr\'echet derivatives, we use a scaling and squaring approach in combination with a Trotter-Suzuki splitting of the elementary propagator. Specifically, we compute the matrix exponent
\begin{align}
e^{X}=\left(e^{{2^{-s}X}}\right)^{2^s}
\label{eq:expTrot}
\end{align}
by $s \in \mathbb{N}$ repeated squaring starting from $e^{{2^{-s}X}}$ calculated as 
\begin{align*}
e^{{2^{-s}X}}= e^{{-i2^{-s-1}}\Delta t\hat{A}'}e^{{-i2^{-s}}\Delta t\hat{A}''}e^{{-i2^{-s-1}}\Delta t\hat{A}'}+O\left(\frac{\Delta t^2}{2^{2s}}\right),
\end{align*}
with 
\begin{align}
X=-i\hat{A}'dt -i\hat{A}''dt
\end{align}
being a convenient splitting of $X=-i dt \hat{A}_i$ (cf. eq.~(\ref{eq:frechet}). Practically, $\hat{A}'$ and $\hat{A}''$ represent the drift and control Hamiltonians, respectively. If the latter is based on coupling applied magnetic fields via the Zeeman interaction, the propagator under $\hat{A}''$ can be expressed as a direct product of evolution operators for the individual electron spins based on expression 
\begin{align}
e^{-i\omega \hat{\rm n}\cdot S}=\cos\left(\frac{\omega}{2}\right)I - 2i\sin\left(\frac{\omega}{2}\right)\hat{\rm n}\cdot S
\label{eq:zeeman}
\end{align}
where $\omega \in \mathbb{N}$ is a parameter related to the strength of the control field and $n$ in a unit vector specifying the orientation of the perturbation in the laboratory frame. 

Likewise, the Fr\'echet derivative of the exponential $e^X$ can be evaluated using an $s$-fold recurrence based on \cite{expFr},
\begin{align}
L(X,E)= e^{X/2} L(X/2, E/2)+L(X/2, E/2)e^{X/2},
\end{align}
obtained by differentiating $e^X=({e^{X/2})}^2$ using the chain rule. In combination with eq.~(\ref{eq:expTrot}), the only derivatives that are explicitly required in this process are those of $e^{{-i2^{-s}}\Delta tA''}$. For Zeeman coupling with a single spin propagator as given by eq.~(\ref{eq:zeeman}), the necessary Fr\'echet derivatives can be obtained in analytic form, in combination with the chain rule, from
\begin{align}
L(\omega \hat{\rm n}\cdot S, -i\hat{\rm n}\cdot S) = -i\cos\left(\frac{\omega}{2}\right)\hat{\rm n}\cdot S - \frac{1}{2}\sin\left(\frac{\omega}{2}\right)\mathbb{1}.
\end{align}
We optimize using the limited memory variant of the Broyden–Fletcher–Goldfarb–Shanno (L-BFGS-B) algorithm, that preconditions gradients using curvature information, and use the SciPy \cite{scipy} implementation with a stopping criterion of $10^{-6}$ and $10^{-7}$ for the change of the sparsely sampled yield and the projected gradient, respectively. The maximum number of iterations was set to 200. We further ran simulations in parallel for varying number of blocks $B$ and the sampling interval $S$, as seen in Figs.~\ref{fig:maeda2}(a) and \ref{fig:maeda2}(b), to assess the performance and scalability of our approach. 

\normalsize

\bibliography{refs.bib}

\clearpage

\onecolumngrid 
\hoffset = 21pt
\voffset = 17pt
\textheight=649pt
\textwidth= 465pt
\fontsize{10}{14}\selectfont 

\setcounter{figure}{0} \newpage
\setcounter{equation}{0}
\renewcommand{\theequation}{S\arabic{equation}}
\renewcommand{\thefigure}{S\arabic{figure}}

\par{\centering
		{{\large \textbf{Supplementary Material}\\ \bigskip\par\vspace{-0.8em}\large{\underline{Quantum Control of Radical Pair Dynamics beyond Time-local Optimization}} } \\ \bigskip\par\vspace{-0.8em}Farhan T. Chowdhury, Matt C. J. Denton, Daniel C. Bonser, and Daniel R.\ Kattnig$^*$ \\ \bigskip\par\vspace{-1.1em}  Department of Physics and Living Systems Institute\\ \bigskip\par\vspace{-1.2em} University of Exeter, Stocker Road, Exeter, Devon, EX4 4QD, United Kingdom\\ \bigskip\par\vspace{-1.2em} \normalsize{E-mail: d.r.kattnig@exeter.ac.uk}}
	\bigskip\par}
 \vspace{-1.5em}
\section*{Additional Results}

We provide further results for the minimization of the recombination yield of the 5-spin system in the low-magnetic field regime, including the geomagnetic field. This is followed by additional results for the 5-spin system controlled under $100\mskip3mu\mu$T, with and without a speed limit imposed. An additional figure shows the frequency profile and recombination yields realized using L-BFGS-B complemented GRAPE for increasing sample intervals for the 18 spin PY/DCB system driven by an RF control field of $0.1\mskip3mu$mT.

\begin{figure} [h]
	\centering
	\includegraphics[scale=0.68]{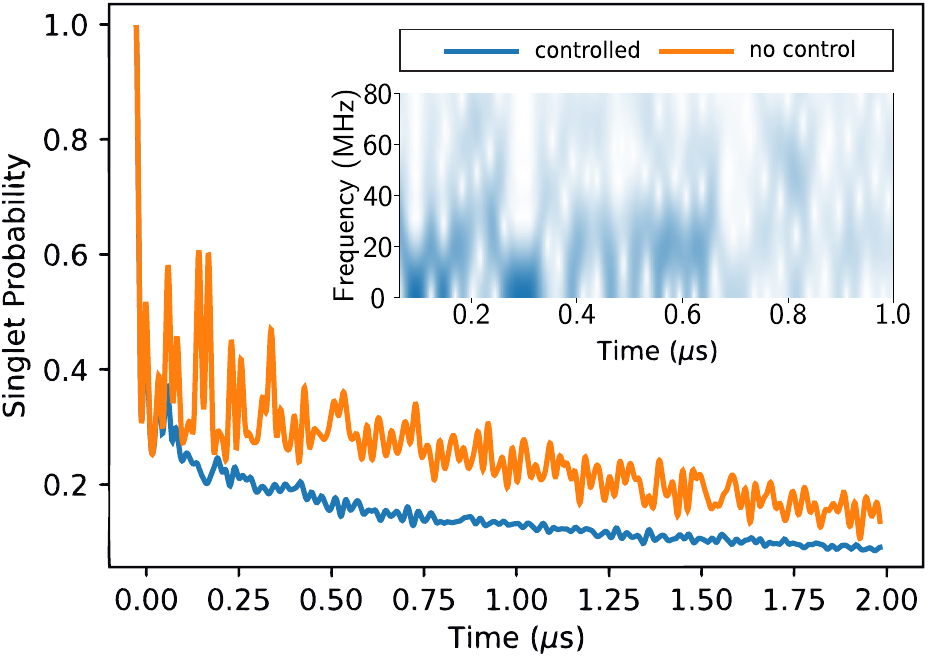}
    \vspace{-1em}
	\caption{Using GRAPE to control the 5-proton spin system in the geomagnetic field ($50\mskip3mu\mu$T), with the singlet probability optimized over 5000 time-steps of $1\mskip3mu$ns using 2 cycles of 2500 time-steps and 250 sampling points. The control amplitude amounted to $0.25\mskip3mu$mT.}
	\label{fig:extras}
    \vspace{-1.5em}
\end{figure}

\begin{figure} [h]
	\centering
	\includegraphics[scale=0.68]{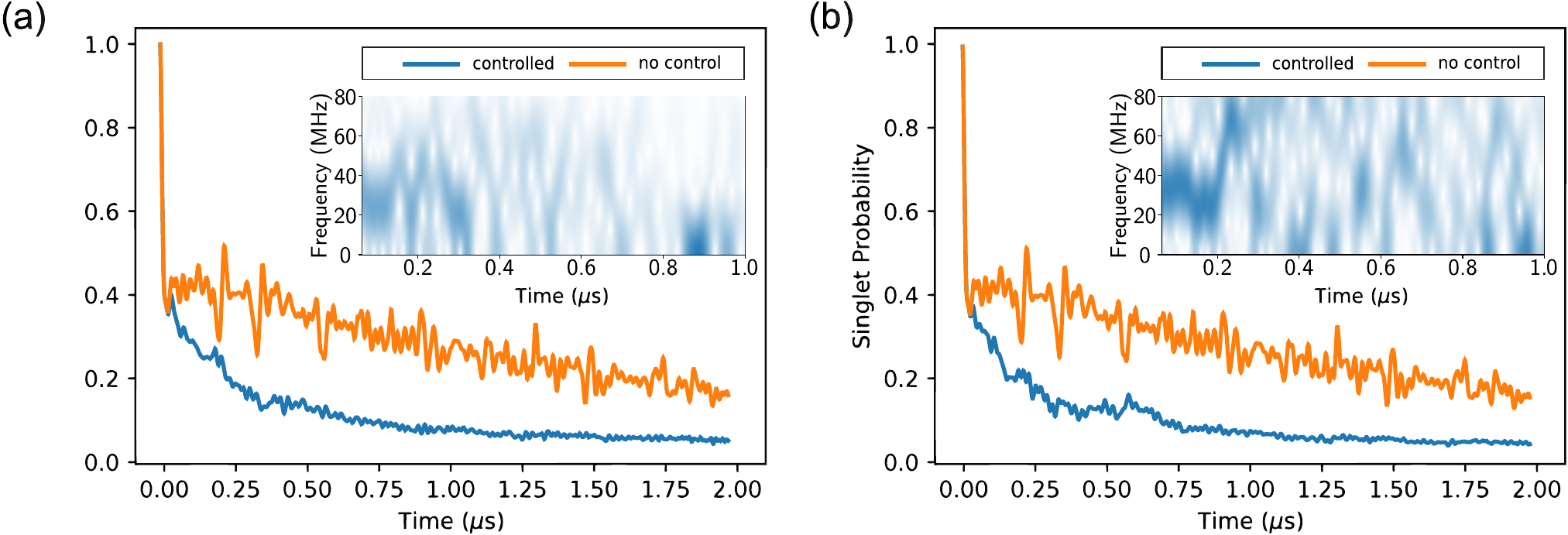}
    \vspace{-1em}
	\caption{Using GRAPE to control the 5-proton spin system under 100$\mu$T magnetic field, subject to a speed limit of $\dot{u}_{max} = 0.2\mskip3mu$ns$^{-1}$ in (a), and without in (b), with the singlet probability optimized using 2 cycles of 2500 time-steps and 250 sampling points.}
	\label{fig:maedaE}
    \vspace{-1em}
\end{figure}

\begin{figure} [b]
	\centering
	\includegraphics[scale=0.48]{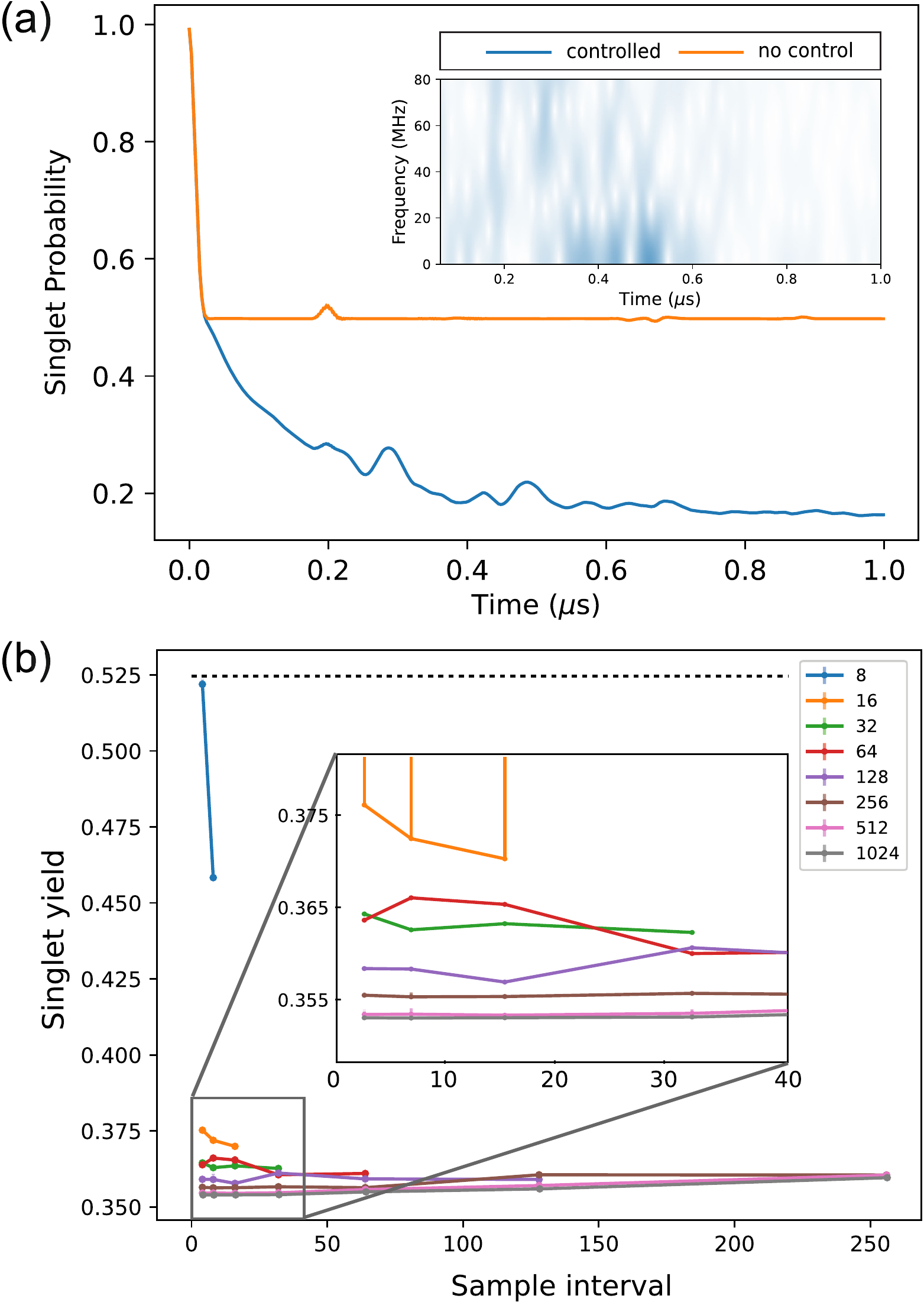}
    \vspace{-1em}
	\caption{We show the frequency profile and recombination yield minimization using L-BFGS-B complemented GRAPE for increasing sample intervals for the 18 spin PY/DCB system, driven by an RF control field of 0.1$\mskip3mu$mT in a static biasing field of $1\mskip3mu$mT using 1024 control pulses of $1\mskip3mu$ns width.}
	\label{fig:pyDCBe}
\end{figure}

\begin{table}[!h]		
\begin{center}	
\caption{Dependence of the optimized recombination yield and runtime of the L-BFGS-B algorithm, as implemented in scipy.optimize.minimize, on the tolerancy ftol. 15 attempts to minimize the recombination yield of the 7-spin model system starting from random initial states uniformly distributed in $[-1,1]$ were recorded. The biasing field amounted to $1\mskip3mu$mT and the maximal control field amplitude to $0.25\mskip3mu$mT. The minimal yield realized was $0.1336$. The iterative algorithm stops when $(f^k -f^{k+1})/max(|f^k|,|f^{k+1}|,1) \leq ftol$, where $f^k$ is the sparse approximation of the yield at iteration $k$. For the application considered here, the denominator equals 1. $N = 5000$, $B = 5$, $S = 200$.}
\vspace{1em}
\begin{tabular}{ |p{1cm}||p{4.5cm}|p{4.5cm}|p{2cm}|  }
 \hline
ftol& Recombination Yield (minimal)& Recombination Yield (average)& runtime (s)\\
 \hline
4   & 0.2665    &0.2738	±	0.0039&75.6   ±	0.8\\
4.5& 0.1354  &0.1496	±	0.0367&1522  ±	480\\
5&0.1345 &0.1376	±	0.0019&2132 ± 298\\
6& 0.1343 &0.1374	±	0.0022&6483 ± 543\\
7& 0.1338  &0.1375	±	0.0022&8886 ± 447\\
8& 0.1336 &0.1372	±	0.0020&9959 ± 113\\
 \hline
\end{tabular}
\end{center}
\vspace{-2.5em}
\end{table}

We explore the question of robustness for our derived controls for two pertinent scenarios, firstly in the case of ``inhomogeneous broadening” due to many (weaker) additional hyperfine interactions, and secondly in the case of spin relaxation as induced by random field fluctuations. As demonstrated in the next section, controls derived for the idealized systems are not invalidated, obviously provided that their effect does not eradicate the magnetic field effects in the absence of control. To further support the case for global optimization over time-local methods, which roughly correspond to what is achievable with GRAPE for insufficient block sizes, we further present plots for 5-spin systems showing singlet recombination yields as a function of $S$ for yield minimizations using $B$ blocks of $N/B$ steps each. The superiority of our approach is contingent upon its ability to capture the time-integrated nature of the reaction yield, the observable quantity, rather than merely the singlet probability at one moment in time. We further provide a quantitative comparison of a generalized version of the time-local algorithm, with required adaptations of the approach described in an additional section, which permits the optimization of the singlet probability, for one instance in time, in the presence of spin selective recombination. 

\begin{figure} [tb]
	\centering
	\includegraphics[scale=0.68]{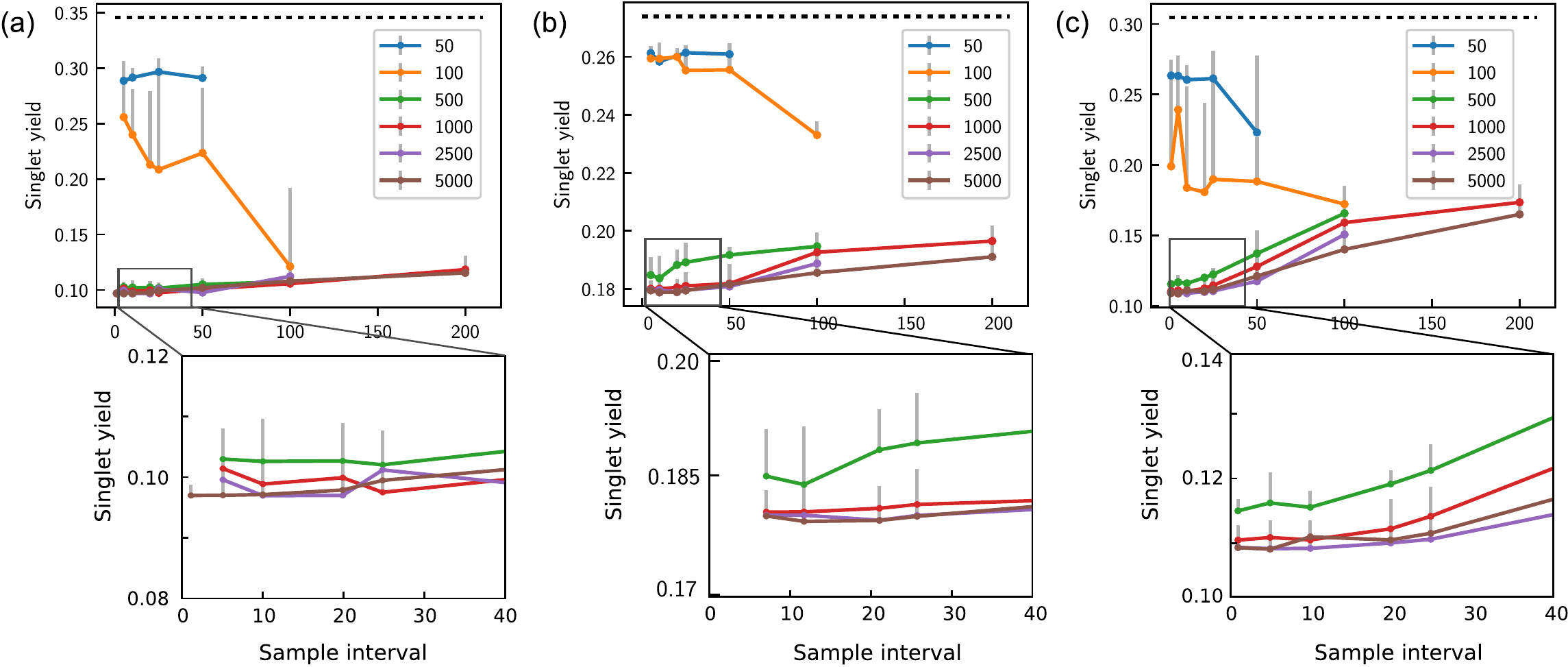}
	\caption{For 5-proton and 3-proton model systems in (a)-(b) and (c), respectively, with a a maximal control field amplitude of $0.25\mskip3mu$mT, with biasing fields of 50$\mskip3mu\mu$T, $10\mskip3mu$mT and $1\mskip3mu$mT for (a), (b) and (c), respectively, we show singlet recombination yield as a function of $S$ for yield minimizations using $B$ blocks of $N/B$ steps each.}
	\label{fig:maedaE2}
\end{figure}

\section*{Robustness to Noise and Unresolved Hyperfine Interactions}

We consider the effects of additional hyperfine interactions, specifically hyperfine interactions that are small compared to the principal hyperfine interactions (which are used when deriving the controls) but numerous, which is typical for the radicals of interest in organic radical recombination reactions. We follow the approach of Schulten and Wolynes \cite{sch_wol78}, applicable for the assumed large number of nuclear spins, according to which the effect of the additional hyperfine interactions can be approximated in terms of additional static random magnetic fields, applied to the two electron spins. 

\begin{figure}[h]
	\centering
	\includegraphics[scale=0.75]{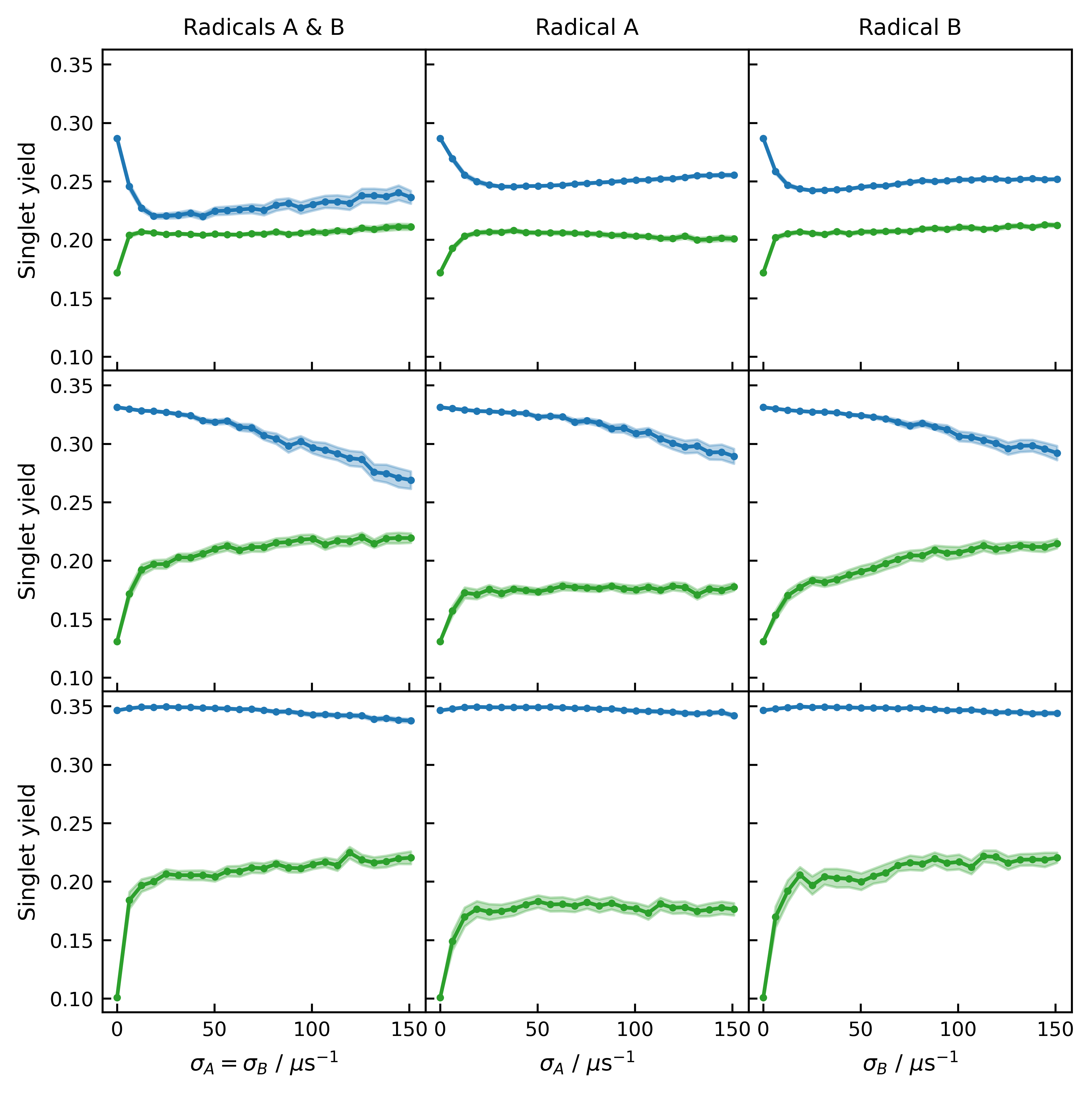}
    \vspace{-1em}
    \caption{Effect of unresolved hyperfine interactions on the singlet recombination yield in the absence (blue) and presence (green) of the control fields for the 7-spin model system studied in the main text. The controls were derived for the idealized systems devoid of inhomogeneous hyperfine interactions, which, if present, were accounted for in the Schulten-Wolynes framework \cite{sch_wol78}. The biasing field amounted to $0\mskip3mu$mT (top), $2\mskip3mu$mT (middle row) and $10\mskip3mu$mT (bottom row); the additional hyperfine fields were added for both radicals (left), or radical A (with 3 dominant hyperfine interactions; central column) or radical B alone (2 dominant hyperfine interactions; right column). The effect of the additional hyperfine interactions was evaluated through Monte Carlo sampling of 100 hyperfine field realizations; the shaded regions correspond to 95$\%$ confidence intervals of the mean recombination yield. The largest added hyperfine fields exceed the intrinsic hyperfine fields of the radicals, thereby demonstrating remarkable robustness of the approach to unresolved hyperfine interactions.}
    \vspace{-1em}
\label{fig:incHyp}
\end{figure}

\begin{figure} [h]
	\centering
	\includegraphics[scale=0.85]{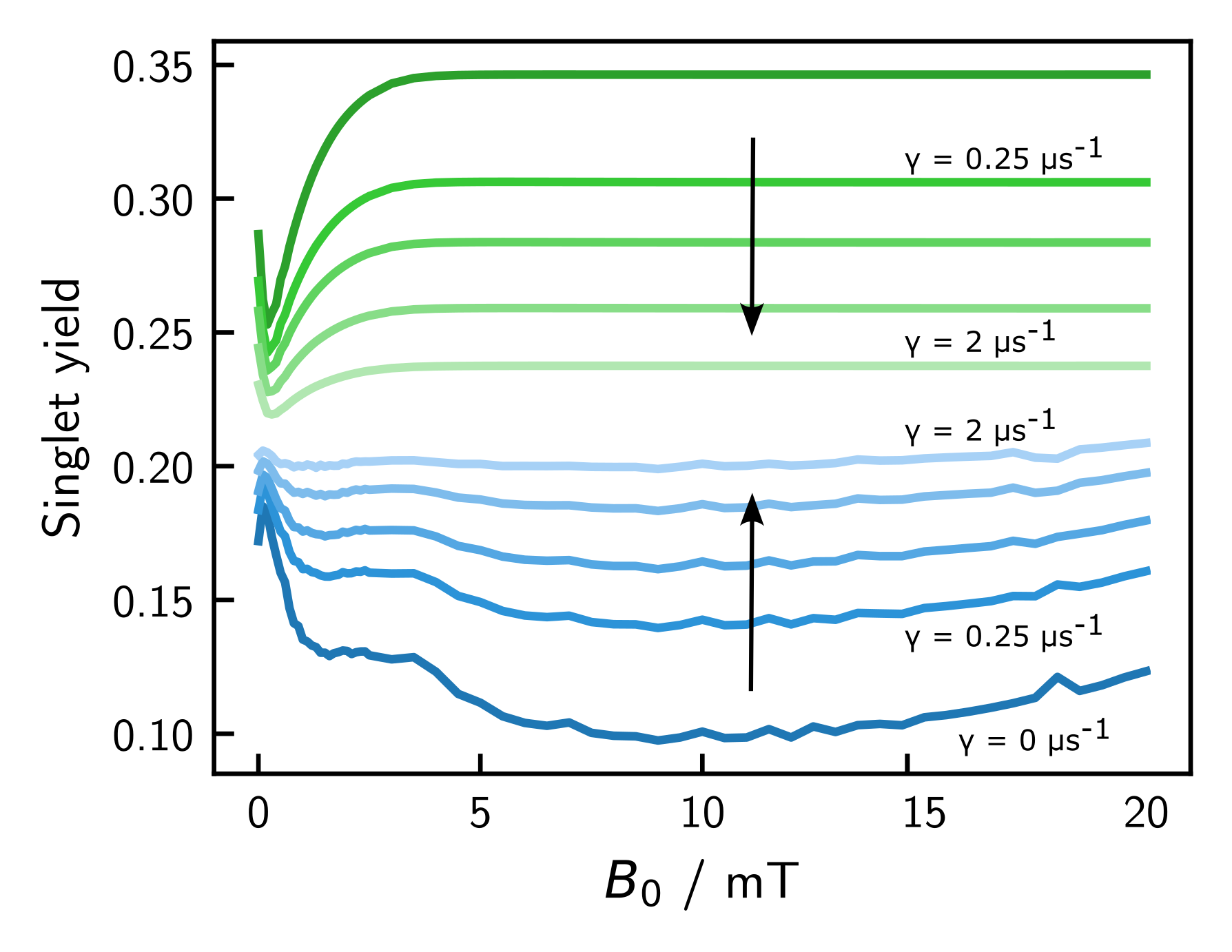}
    \vspace{-1.5em}
    \caption{Dependence of the singlet recombination yield of the 7-spin model spin system studied in Fig.~4 of the main text on the rate of random-field relaxation, $\gamma$, in the absence (green) and presence (blue) of controls. Controls were derived for the idealized system without spin relaxation. $\gamma$ increases with increasing lightness of the curves, comprising the values $0$, $0.25$, $0.5$, $1$, $2\mskip3mu\mu$s$^{-1}$. The latter exceeds the radical pair lifetime ($k_S = k_f = 1\mskip3mu\mu$s$^{-1}$), thus strongly attenuating the magnetic field sensitivity. Yet, the controls remain functional and lead to reaction yields that cannot be realized using static magnetic field of any flux density.}
	\label{fig:spinRel}
\end{figure}

These additional fields comprise identically and independently normally distributed field components of mean 0 and variance $\sigma_i^2 = \frac{1}{3} \sum_j a_{i,j}^2 I_{i,j} (I_{i,j} + 1)$, where the sum runs over all additional hyperfine interactions of radical $i$. Fig.~\ref{fig:incHyp} illustrate the effect of such unresolved hyperfine on the singlet recombination yield for the 7-spin model system, based on Monte Carlo sampling hyperfine fields according to the above paradigm. The controls have been derived based on the idealized system devoid of the additional hyperfine fields and their resilience tested in their presence. We investigated three different biasing field strengths, $0$, $2$ and $10\mskip3mu$mT, for which we increased $\sigma_i$ for one or both radicals up to $150\mskip3mu$rad/$\mu$s. The maximal hyperfine field exceeds the hyperfine field of the dominant interactions, which amounts to $100$ and $32\mskip3mu$rad/$\mu$s (but realized in terms of 3 or 2 hyperfine interactions only). As is apparent from Fig.~\ref{fig:incHyp}, the controls remain robust even in the presence of strong inhomogeneous broadening, realizing recombination yields that cannot be attained through application of static fields alone. To assess the effects of the presence of spin relaxation, we study how noise, modeled through the Lindblad form
\begin{equation}
\hat{\hat{R}}\hat{\rho}(t) = \sum_i \gamma_i \left[ \hat{L}_i \hat{\rho}(t) \hat{L}_i^\dagger - \frac{1}{2} \left\{ \hat{L}_i^\dagger \hat{L}_i, \hat{\rho}(t) \right\} \right]
\end{equation}
with  $\hat{L}_i \in \left\{ \hat{S}_{1x}, \hat{S}_{1y}, \hat{S}_{1z}, \hat{S}_{2x}, \hat{S}_{2y}, \hat{S}_{2z} \right\}$, impact the control success when added to the equation of motion in the main text. Assuming a common relaxation rate for all collapse operators $\hat{L}_i$ (isotropic random field noise), Fig.~\ref{fig:spinRel} shows the effect of increasing the relaxation rate $\gamma$ on the singlet recombination yield for the controlled and unperturbed 7-spin system (cf.\ Fig.~4 in the main text). We chose relaxation rates ranging up to $2\mskip3mu\mu$s$^{-1}$, which exceeds the radical pair lifetime, and consequently strongly reduces the intrinsic magnetic

\clearpage

\noindent field sensitivity, i.e.\ the curves describing the recombination yield as a function of the biasing fields are flattened and the low-field effect attenuated. Nonetheless, controls derived for the idealized non-relaxing system remain effective in reducing the recombination yield beyond the yields realizable through applying static magnetic fields, i.e.\ the controls are robust. 

\section{Comparisons to Time-Local Optimization}

We generalize the approach of \cite{mae2020} to account for spin-selective recombination and observables that do not commute with the effective Hamiltonian. We limit the exposition to the essential facts; the reader is referred to the original publication for details. The spin density operator of a radical pair subject to coherent evolution under the drift Hamiltonian $H_0$ and a single control field with Hamiltonian $H_1$ and amplitude $u(t)$ obeys the following equation of motion in the Liouville representation
\begin{equation}
\frac{d}{dt}|\rho(t)\rangle = -i\left(\hat{L}_0 + u(t)\hat{L}_1 - i\hat{K}\right)|\rho(t)\rangle = -i\left(\hat{L}+ u(t)\hat{L}_1\right)|\rho(t)\rangle,
\end{equation}
where $L_0$ and $L_1$ are commutation superoperators, i.e.\ $\hat{L}_i |\rho\rangle = |[H_i, \rho]\rangle$, $\hat{K}$ accounts for recombination, i.e.\ $\hat{K} |\rho\rangle = |\{K,\rho\}\rangle$, and $\hat{L} = \hat{L}_0 - i\hat{K}$ was introduced in the final equality. We use single hats for super-operators and omit them for usual Hilbert space operators, which induce the ket-states of the Lioville space, denoted by the usual Dirac kets. To optimize the observable $\Omega$ at time $t_f$, we introduce a moving target
\begin{equation}
|\Omega(t)\rangle = \exp(i\hat{L}^\dagger(t_f - t))|\Omega\rangle,
\label{eq:movT}
\end{equation}
where the adjoint operation must be applied as $L$ is not anti-Hermitian as a consequence of the presence of $\hat{K}$ accounting for the radical recombination. With the target state redefined as above, we follow \cite{mae2020} and introduce the optimization quantity
\begin{equation}
y = \langle \Omega(t) | \rho(t) \rangle,
\end{equation}
which for $t=t_f$ equals the expectation value of observable $\Omega$. The time evolution of $y$ is found as
\begin{equation}
\frac{d}{dt}y(t) = -iu(t)\langle \Omega(t) | \hat{L}_1 | \rho(t) \rangle.
\end{equation}
Hence, choosing $u(t)$ to obey
\begin{equation}
u(t) = i A \langle \Omega(t) | \hat{L}_1 | \rho(t) \rangle^{*} \in \mathbb{R}
\label{eq:A_mod}
\end{equation}
for any real constant $A>0$, guarantees that
\begin{equation}
\frac{d}{dt}y(t) = A|\langle \Omega(t) | \hat{L}_1 | \rho(t) \rangle|^2 \geq 0.
\end{equation}
Consequently, $y(t)$ will monotonously increase over time, thereby enabling a (constraint) maximization of $y(t_f)$. The optimization is constrained in the sense that solutions for which $y(t)$ is non-monotonic are excluded. Note furthermore that for $\rho(t=0)\propto P_S$ and $H_1\propto S_x$, $u(t=0)=0$. For $A<0$, a minimization will be realized. Apart from the redefinition of the target state, the approach is identical to \cite{mae2020}, to which we refer for additional details. 
\begin{table}[!h]
\begin{center}
\caption{Recombination yields of the 7-spin radical pair for $B_0=0\mskip3mu$mT subject to the optimal controls as derived using the generalized time-local approach targeting the singlet probability at $t_f$. $A$ is the amplitude parameter (cf.\ eq.~(\ref{eq:A_mod}) and $S$ is the relative success expressed as the change of yield induced by the controls relative to the change of yields realized via the GRAPE-derived controls from the main text.}
\vspace{1em}
\begin{tabular}{ |p{2cm}|p{2cm}|p{2cm}|p{2cm}|p{2cm}|  }
 \hline
$t_f$ ($\mu$s)& A& $Y_S$& S(\%)& $\max |u(t)|$ \\
 \hline
1& -1.48& 0.283& 3.3& 1.0\\
 \hline
1& -1.5& 0.279& 6.7& 1.5\\
 \hline
1& -2& 0.258& 25& 2.2\\
 \hline
1& -2.5& 0.247& 35& 3.0\\
 \hline
1& -3& 0.249& 33& 5.6\\
 \hline
2& -1.5& 0.265& 19& 1.2\\
 \hline
2& -2.5& 0.242& 39& 2.3\\
 \hline
3& -2.5& 0.279& 6.5& 0.9\\
 \hline
3& -2.7& 0.279& 6.8& 1.0\\
 \hline
4& -2& 0.282& 4.5& 1.1\\
 \hline
4& -2.5& 0.277& 8.4& 1.4\\
 \hline
5& -2.5& 0.287& 0.0& 0.0\\
 \hline
5& -100& 0.286& 0.4& 2.1\\
 \hline
\end{tabular}
\end{center}
\vspace{-2.5em}
\end{table}
We implement the time-local optimization approach using the zvode ODE-solver in SciPy \cite{scipyS}, which employs the Adams/BDF-method of order 12. We consider the problem of minimizing the singlet recombination yield of the 7-spin model system, proxied through the singlet probability at time $t_f$. For no biasing field ($B_0 = 0$), our GRAPE-inspired approach yielded a reduction of the recombination yield from $0.287$ to $0.172$ using controls of maximal amplitude $0.25\mskip3mu$mT (corresponding to $u(t)=\pm1$) when controlling the first $5\mskip3mu\mu$s, with controls established through optimization of 5 blocks of 1000 control amplitudes each, and sparse sampling of the yield. 

\begin{figure} [h]
	\centering
	\includegraphics[scale=0.9]{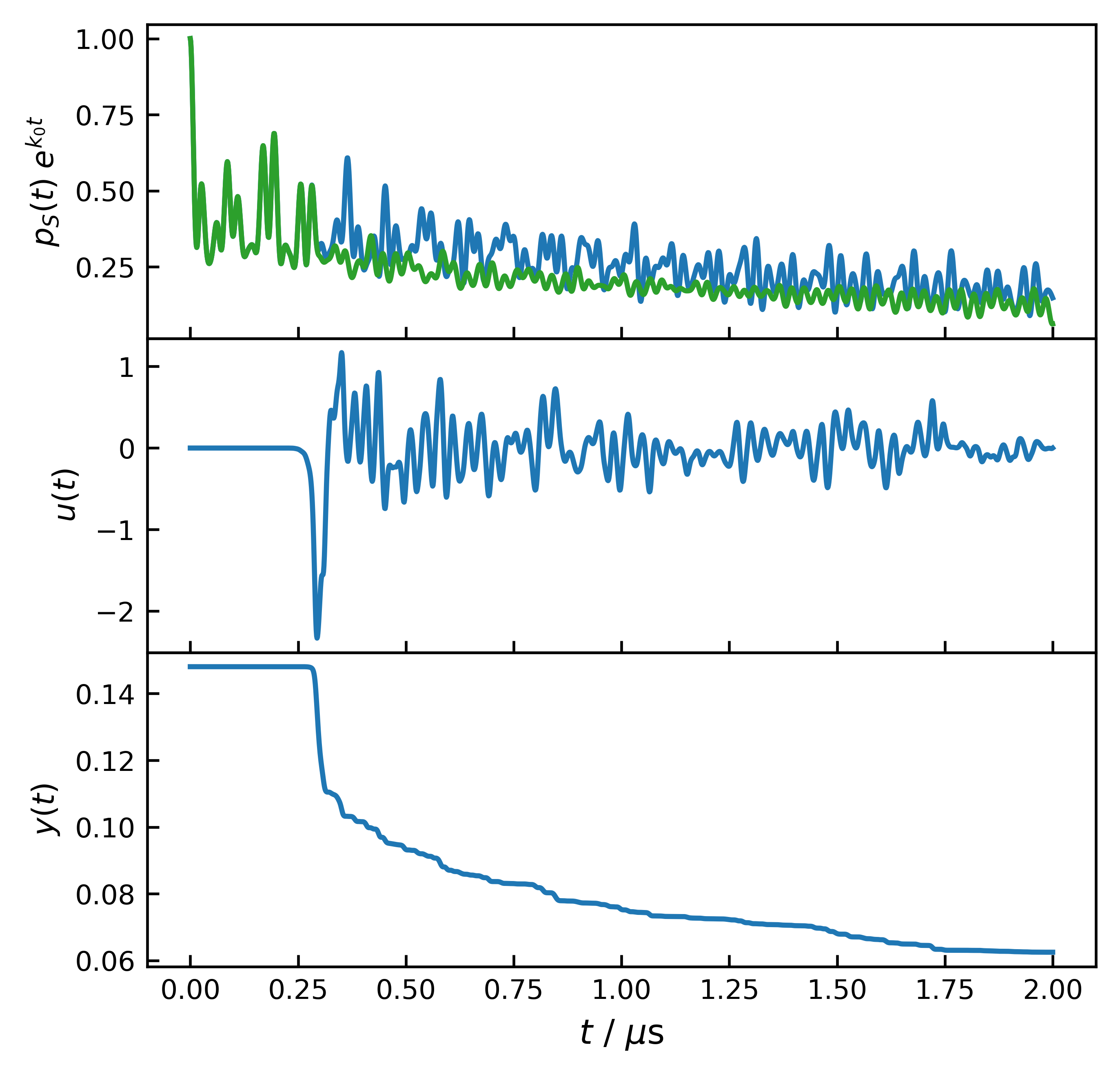}
    \vspace{-1.5em}
    \caption{Time-local optimization of the 7-spin model system studied in the main text. The optimization was realized for $B_0 = 0\mskip3mu$mT through a generalization of the approach from \cite{mae2020}, detailed above, with $A = -2.5$. The moving target (cf.\ eq.~(\ref{eq:movT}) optimized the singlet probability at time $t_f = 2\mskip3mu\mu$s. The singlet recombination yield was reduced from $0.287$ to $0.242$ using a maximal control amplitude $\max |u(t)|\approx2.3$ (the GRAPE-inspired approach yielded superior minimization with $Y_S = 0.172$). The panels give the singlet probability, rescaled to remove the effect of the spin-independent reaction, the relative control amplitude (multiplying the nominal amplitude of $0.25\mskip3mu$mT), and the optimization criterion $y(t)$, which evolves monotonously in time.}
\end{figure}

\clearpage

First, we applied the local optimization approach for $t_f$ in the range from $1$ to $5\mskip3mu\mu$s, choosing the amplitude parameter $A$ (cf.\ eq.~(\ref{eq:A_mod}) such that a comparable maximal control amplitude of $\max |u(t)|\approx1$ resulted, and as $A = -2.5$, which yielded larger control amplitudes for most target times $t_f$. Furthermore, for $t_f = 1\mskip3mu\mu$s, additional $A$-parameters up to $|A| = 3$ were tried. Inspection of the control results, summarized in Table.~(II), leads to the quantitative insight that the time-local control approach is inferior for all $t_f$. When limited to the approximately same maximal control amplitude, the ``best'' reduction of the recombination yields realized did not permit $Y_S$ below 0.265. Better outcomes could only be achieved using (markedly) larger control amplitudes. For example, for $t_f = 1\mskip3mu\mu$s, for $A = -2.5$ we obtained $Y_S = 0.247$, realized using a control amplitude that exceeded those used in our approach by nearly a factor of 3. For $t_f = 2\mskip3mu\mu$s, $A = -2.5$ yielded $Y_S = 0.242$ at more than twice the intended control amplitude.

We note that further increasing $|A|$ did not allow approaching the GRAPE-based results, as is apparent from the data for $t_f = 1\mskip3mu\mu$s in Table.~(II). Even for the 5.6-fold increased maximal amplitude realized for $A = -3$, the yield is limited to $0.249$. Fig.~(S7) illustrates the optimization approach for $t_f = 2\mskip3mu\mu$s and $A = -2.5$. Note that the optimized $u(t)$ is approximately zero for the initial time period, i.e.\ a phase of unperturbed evolution precedes the control part, which is a characteristic feature observed for all time-local optimizations realized here and also seen in the test cases from \cite{mae2020}. Given that controls start out at zero and observing that overall better results are obtained for the shorter $t_f$ tried, we wondered if our blocking approach successfully applied for the new algorithm, as laid out in the main text, could likewise enhance the optimization of the time-local approach, as discussed here. To this end, we focused on the first $5\mskip3mu\mu$s of the evolution, as for the GRAPE-inspired approach, and subdivided this interval in non-overlapping blocks. The optimization was carried out by re-defining the moving target for every block such that it yielded the singlet probability at the respective block end time. Indeed, in this way, we were able to realize an improvement over the results for a single $t_f$. However, again, larger amplitudes were required, and the results still fell short of the GRAPE-based approach. Table.~(III) summarizes the optimization results for $A = -2.5$, which gives rise to amplitudes markedly exceeding $|u|=1$, for the smallest block size by a factor of 23. The best result was obtained for 80 blocks, giving $Y_S = 0.208$ with $\max |u(t)|\approx8.9$. While this yield still significantly overshoots the GRAPE-based $Y_S = 0.172$, we think that time-blocking might be an interesting option for the time-local optimization approach.

\begin{table}[!h]
\begin{center}
\caption{Recombination yields of the 7-spin radical pair for $B_0=0\mskip3mu$mT subject to the optimal controls as derived using the time-local approach targeting the singlet probability for consecutive time blocks. The first $5\mskip3mu\mu$s of the time evolution where controlled with controls optimized for blocks of length $t_b$. $A$ is the amplitude parameter (cf.\ eq.~(\ref{eq:A_mod}) and $S$ is the relative success expressed as the change of yield induced by the controls relative to the change of yields realized via the GRAPE-derived controls from the main text.}
\vspace{1em}
\begin{tabular}{ |p{2cm}||p{2cm}|p{2cm}|p{2cm}|p{2cm}|p{2cm}|  }
 \hline
No.\ of blocks& $t_b$ ($\mu$s)& A& $Y_S$& S (\%)& $\max |u(t)|$ \\
 \hline
5& 1& -1.48& 0.272& 13& 1.9\\
 \hline
5& 1& -2& 0.258& 25& 2.2\\
 \hline
5& 1& -2.5& 0.248& 33& 3.0\\
 \hline
5& 1& -3& 0.248& 34& 5.6\\
 \hline
10& 0.5& -2.5& 0.239& 42& 3.3\\
 \hline
20& 0.25& -2.5& 0.214& 64& 8.4\\
 \hline
40& 0.125& -2.5& 0.228& 51& 8.2\\
 \hline
80& 0.0625& -2.5& 0.208& 69& 8.9\\
 \hline
160& 0.03125& -2.5& 0.208& 69& 14.7\\
 \hline
320& 0.015625& -2.5& 0.215& 62& 21.1\\
 \hline
\end{tabular}
\end{center}
\label{fig:tB}
\end{table}
 
\begin{figure} [h]
	\centering
	\includegraphics[scale=0.9]{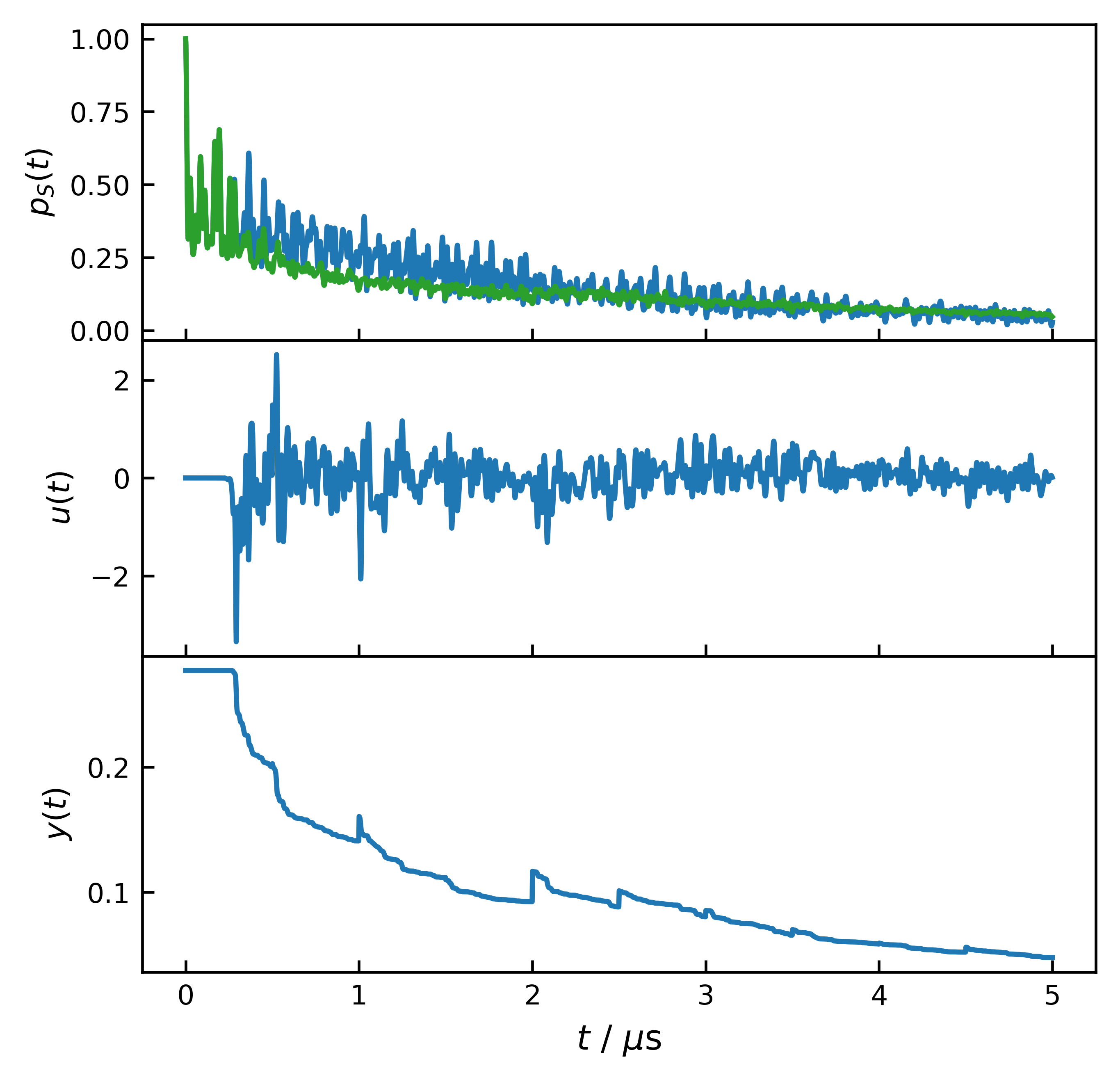}
    \vspace{-1.5em}
    \caption{Time-blocked time-local optimization of the 7-spin model system studied in the main text. The optimization was realized for $B_0 = 0\mskip3mu$mT through a generalization of the approach from Ref.\ \cite{mae2020}, detailed in the appendix, with $A = -2.5$, and application of a time-blocking approach like that used here for the GRAPE-inspired optimization. The first $5\mskip3mu\mu$s of evolution were subdivided in 10 blocks, to which the time-local optimization with moving target optimizing the singlet probability at the blocks’ ends was applied. The redefinitions of the targets give rise to jumps in $y(t)$, which otherwise decreases monotonously, as enforced by the time-local approach. The singlet recombination yield was reduced from $0.287$ to $0.239$ using a maximal control amplitude (the GRAPE-inspired approach yielded superior minimization with $Y_S = 0.172$). Using 20 to 160 blocks gave rise to better optimization results (cf. Table.~III), but the characteristic features of $y(t)$ are less apparent.}
\end{figure}

\end{document}